\pdfoutput=1
\documentclass[a4paper,11pt]{article}
\usepackage{graphicx}
\usepackage{caption}
\usepackage{subcaption}
\usepackage{jheppub-mod}

\usepackage[utf8]{inputenc}
\usepackage[T1]{fontenc}
\usepackage{amsmath}
\usepackage{amsfonts}
\usepackage{amssymb}
\usepackage{bbm}
\usepackage{braket}
\usepackage{xcolor}

\graphicspath{{plots/}}

\usepackage{ulem}

\title{Kinematic enhancement for nucleon interpolators}

\preprint{MIT-CTP/6044}

\author[a]{Daniel Reitinger,}
\author[a]{Tobias Sizmann,}
\author[a]{Andreas Sch{\"a}fer,}
\author[b]{Rui Zhang,}
\author[c]{Yong Zhao}

\affiliation[a]{Institute for Theoretical Physics, University of Regensburg,
   93040 Regensburg, Germany}
\affiliation[b]{Center for Theoretical Physics - a Leinweber Institute, Massachusetts Institute of Technology, Cambridge, MA 02139, USA}
\affiliation[c]{Physics Division, Argonne National Laboratory, Lemont, IL 60439, USA}

\abstract{The reliable treatment of highly boosted hadrons is crucial for many lattice QCD applications. For all these cases the kinematically-enhanced interpolators advocated in \cite{Zhang:2025hyo} promise very significant improvements and are, therefore, ever more often used in recent calculations, especially for highly boosted mesons like the pion. 
Motivated by, e.g., the physics program of the future Electron-Ion Collider (EIC) in the US and Electron-Ion collider in China (EIcC), we systematically benchmark our code for the unpolarized isovector nucleon quark matrix elements extracted at large source-sink separations, where excited state artifacts are significantly suppressed. We find that the precision of the renormalized nucleon matrix elements is typically improved by an order of magnitude at momentum $P_z\sim2.5$ GeV.
By comparing the results from three CLS ensembles with different lattice spacings $a$ but the same pion mass, we observe no statistically significant dependence on $a$ in the renormalized matrix elements at nearly identical values of $P_z$. These encouraging results suggest that the use of kinematically improved operators is highly advantageous for parton physics calculations and can be extended to a broader class of baryon observables, making them a promising candidate for a standard component of modern lattice QCD.}

\emailAdd{Daniel.Reitinger@physik.uni-regensburg.de}

\begin{document}

\maketitle


\section{Introduction}
In recent years several methods~\cite{Liu:1993cv,Detmold:2005gg,Braun:2007wv,Davoudi:2012ya,Ji:2013dva,Radyushkin:2017cyf,Ma:2017pxb,Chambers:2017dov,Shindler:2023xpd} have been proposed and widely applied to extract far more information on the partonic structure of hadrons from lattice simulations than was judged possible before. To a large extent, this effort is motivated by new and more powerful and versatile accelerator facilities like JLab, LHC, EIC, and EIcC. We do not attempt to provide a comprehensive overview of these experimental and theoretical efforts, for which substantial review literature already exists, such as Refs.~\cite{Anderle:2021wcy,Burkert:2022hjz,Amoroso:2022eow,Boussarie:2023izj,Accardi:2023chb,Abir:2023fpo,Constantinou:2020pek,Ji:2020ect}. However, we emphasize the importance of achieving good signal-to-noise ratios (SNRs) in order to fully exploit the significantly broadened spectrum of physical effects. Many proposed theoretical approaches require highly boosted hadrons to access the full kinematic information of parton distributions, including Large Momentum Effective Theory (LaMET)~\cite{Ji:2013dva,Ji:2014gla,Ji:2020ect}, pseudo-distributions~\cite{Radyushkin:2017cyf}, current-current correlators~\cite{Braun:2007wv,Ma:2017pxb,Zhang:2026lle}, and hadronic tensor~\cite{Liu:1993cv} or Compton amplitudes~\cite{Detmold:2005gg,Chambers:2017dov}. Highly boosted hadrons are also crucial for studying form factors at large momentum transfer~\cite{Ding:2024lfj}, heavy-meson and baryon decays~\cite{LHCb:2019hro,LHCb:2025ray}, and many other physical processes.

With this in mind, we focus on LaMET, the most widely used framework for parton physics calculations so far~\cite{Cichy:2018mum,Zhao:2018fyu,Lin:2017snn,Constantinou:2020pek,Constantinou:2020hdm,Ji:2020ect,Cichy:2021lih,Zhao:2025oto,Lin:2025hka,Proceedings:2026xrb}, 
as we expect our conclusions to remain qualitatively applicable to other methods and physical problems. LaMET relies crucially on an expansion of spatial quasi-observables around the limit $P_z\to\infty$ to calculate the $x$-dependence of corresponding light-cone partonic observables~\cite{Ji:2024oka,Ji:2020byp,Ji:2022ezo}, including the parton distribution functions (PDFs), Generalized Parton Distributions, Transverse Momentum Distributions, and Wigner distributions. Taking this limit is, however, highly nontrivial due to the rapidly deteriorating SNRs of hadronic observables with increasing $P_z$. For the nucleon, the experimentally most relevant hadron, this typically makes it difficult to reach momenta beyond $P_z=3$~GeV on contemporary lattice ensembles. This precision bottleneck also affects the calculation of quasi-distributions through the Fourier transformation of nonlocal hadronic matrix elements, a topic that has attracted considerable recent discussion~\cite{Dutrieux:2025jed,Xiong:2025obq,Chen:2025cxr,Ling:2025olz,Medrano:2025cmg,Ji:2026vir}.

Ref.~\cite{Zhang:2025hyo} introduced a methodological improvement that enhances the SNRs of highly boosted hadrons on the lattice, which has been applied to recent calculations~\cite{Bollweg:2025iol,LatticePartonCollaborationLPC:2025vhd,Detmold:2025lyb,ChenChen:2025amm,Tan:2025ofx,CLQCD:2025dod,LPC:2026vyv}.
While these publications primarily presented results for two-point functions and effective energies, we investigate the improvement for nucleon quasi-distribution matrix elements. We implement, optimize, and fine-tune the algorithms for CLS ensembles, the physics program of the Lattice Parton Collaboration (LPC), and GPU architectures. 
To quantify the improvement, we compute matrix elements of the unpolarized isovector nucleon quark quasi-parton distribution function (quasi-PDF) in a gauge-invariant formulation. We furthermore investigate the results for possible increased excited-state overlap as well as discretization and cutoff artifacts. To this end, we make a direct comparison of quasi-PDF matrix elements obtained using conventional and kinematically enhanced interpolators at different lattice spacings at similar momenta. Our results indicate that the improved interpolators enhance the statistical precision of boosted nucleon matrix elements at $P_z\sim 2.5$ GeV by more than an order of magnitude, suggesting that significantly more ambitious simulations should be feasible. This significant improvement is also partly due to the projector $\gamma_+$ contained in the enhanced interpolator, which yields an additional factor-of-four reduction in variance. One factor of two arises from combining the information contained in the $\gamma_t$ and $\gamma_z$ matrix elements, while the other originates from $\gamma_+$ eliminating all ``$-$''-components of the propagators, thereby reducing the computational cost of their evaluation. Besides, the quasi-PDF matrix elements renormalized in the hybrid scheme~\cite{Ji:2020brr} exhibit a well-behaved continuum limit, indicating that the discretization effects remain small even at momenta as large as $P_z\sim 2.5$~GeV. Finally, we also observe that the SNR enhancement increases with finer lattice spacing.

This paper is organized as follows. In Section~\ref{section:theory}, we review the theory of enhanced interpolators and outline the basics of PDF calculations within LaMET. Section~\ref{section:setup} presents the lattice setup used in our simulations, while Section~\ref{section:results} contains our numerical results, demonstrating that kinematically enhanced interpolators are highly efficient for nucleons. Finally, we conclude in Section~\ref{section:summary}.

\section{Theory}
\label{section:theory}
\subsection{Kinematic enhancement}
In lattice QCD, the hadrons are simulated with interpolating operators built from the quark and gluon degrees of freedom, i.e., Fock states constructed of quark and gluon fields. A symmetric two-point correlation function admits a spectral decomposition,
\begin{align}
    C_{\rm 2pt}(t)=\langle \mathcal{J}^\dagger(t)\mathcal{J}(0)\rangle=\sum_i \frac{|Z_i|^2}{2E_i}e^{-E_i t},
\end{align}
where the overlap factor $Z_i = \langle i| \mathcal{J}|\Omega\rangle$ depends on the wave function of the hadron $|i\rangle$ projected onto the corresponding Fock state $\mathcal{J}|\Omega\rangle$. To have non-vanishing overlap, the interpolating operator $\mathcal{J}$ must have the same quantum numbers and transform in the same irreducible representation as the physical state $|i\rangle$. In a boosted system, the lattice symmetry group is reduced from the cubic group $O_h^D$ ($O_h$) to the little group ${\rm Dic}_{4}$ ($C_{4v}$) corresponding to the 4-fold rotation around the longitudinal direction\footnote{Note that the little group will be different if the hadron's momentum is not pointing in direction of an axis, but this does not affect the following discussion.}. Thus the hadron has non-vanishing overlaps with a larger set of Fock states.  Among them, the leading component at large momentum, or near the lightcone, is described by the leading-twist wave function of the hadron~\cite{Burkardt:2002uc,Ji:2003yj}. This is related to the fact that the ``$+$'' component $\psi_+=\frac{1}{\sqrt{2}}\gamma_-\gamma_+\psi$ is the only dynamic degree of freedom on the lightcone. Here $\gamma_{\pm}=\frac{1}{\sqrt{2}}(\gamma^M_t\mp\gamma^M_z)$ (assuming the hadron is moving in the $\hat{z}$ direction) in Minkowski spacetime, which is related to the Euclidean definition through $\gamma_4=\gamma^M_t$ and $\gamma_{\{1,2,3\}}=\mathbf{i}\gamma^M_{\{x,y,z\}}$. On a Euclidean lattice, although the lightcone does not exist, the ``$+$'' components still dominate at large but finite momentum in the form of a kinematical enhancement~\cite{Zhang:2025hyo}. For a leading-twist interpolator $J_+$ constructed with $n$ ``$+$''-component quarks, the overlap with the hadron $H(P)$ moving at momentum $P$ is kinematically enhanced as,
\begin{align}
    \langle0|{\mathcal{J}_+}|H({P})\rangle \propto P_+^{n/2},
\end{align}
where $P_+=(E+P_z)/\sqrt{2}$.
As a result, the amplitude of the corresponding two-point correlation function is enhanced as
\begin{align}
    |Z_0|^2=|\langle0|{\mathcal{J}_+}|H({P})\rangle|^2 \propto P_+^n.
\end{align}

The fluctuation of the measurement can be estimated by the variance operator~\cite{Parisi:1983ae,Lepage:1989hd},
\begin{align}
    {\rm Var}(C_{\rm 2pt})&=\frac{1}{2}\langle|C_{\rm 2pt}|^2\rangle + \frac{1}{2}\langle C_{\rm 2pt}^2\rangle - \langle C_{\rm 2pt}\rangle^2,\nonumber\\
    &=\frac{1}{2}\langle \mathcal{J}^\dagger(t)\mathcal{J}(t)\mathcal{J}^\dagger(0)\mathcal{J}(0) \rangle + \frac{1}{2}\langle \mathcal{J}^\dagger(t)\mathcal{J}^\dagger(t)\mathcal{J}(0)\mathcal{J}(0)\rangle - \langle \mathcal{J}^\dagger(t)\mathcal{J}(0)\rangle^2,
\end{align}
where the first term is the two-point correlation for a static or scattering state with zero total quantum number and momentum, and the second term is for a two-particle state with momentum $2P$. At large Euclidean time, the variance is dominated by the lowest-energy state, corresponding to $n$ static pions for $n$ valence quarks. Thus the fluctuation of the correlator does not get a kinematic enhancement, resulting in an improved signal-to-noise ratio,
\begin{align}
    \frac{C_{\rm 2pt}(t\to\infty)[\mathcal{J}_+]}{\sqrt{{\rm Var}[C_{\rm 2pt}(t\to\infty)]}}\propto P_+^ne^{E_0-nm_{\pi}/2}.
\end{align}
In comparison with the static system, except for the exponential decay, the signal-to-noise ratio from the overlap factor $|Z_0|^2$ is enhanced by the factor $\frac{P_+^n}{M^n}$. Since $P_+\propto E+P_z$, both the $\mathcal{J}_4$ and $\mathcal{J}_{\hat{P}}$ components get enhanced by the factor $\frac{E^n}{M^n}$ and $\frac{P_z^n}{M^n}$, respectively. Note that the $\mathcal{J}_4$ component has non-vanishing overlap with the ground-state hadron at rest, and thus has already been well studied for static hadrons in the literature. 

\subsection{Interpolator kernels}
In this subsection we present the diquark and free-quark kernels we use in this study, with the following proton interpolators~\cite{Zhang:2025hyo}
\begin{align}
    \label{eq:Interpolators}
    &{\mathcal{J}} = \mathcal{P}\epsilon_{abc}{u}_a\left( {u}_b\Gamma^\prime {d}^T_c \right) \;, \nonumber \\ 
    &\overline{\mathcal{J}} = \epsilon_{abc}\bar{u}_a \gamma_4 \mathcal{P}^\dagger \gamma_4 \left( \bar{u}_b\gamma_4\Gamma^{\prime *}\gamma_4 \bar{d}^T_c \right) \;,
\end{align}
where $\mathcal{P}$ projects onto the desired degrees of
freedom.  $\Gamma'$ is the diquark kernel.

Both the two-point and the three-point function include a contraction of the respective unpaired quarks in the source and sink interpolators. We summarize the contraction of the projectors as $\mathcal{T}$, the free-quark kernel. The two-point function $C_{2pt}$ reads 
\begin{equation}
\small
\begin{aligned}
C_{2pt}(\mathbf{p},t)
&= \langle 0 | \mathcal{J}(\mathbf{p},t)\, \overline{\mathcal{J}}(\mathbf{p},0) | 0 \rangle \\
&=
a^3\sum_{\mathbf{x}}
e^{-i\mathbf{p}\cdot\mathbf{x}}\;
\epsilon_{abc}\,\epsilon_{a'b'c'}\;
\\
&\quad\times
\Big\langle 0 \Big|
\text{Tr}\left[\bar{u}_{a'}(0)\,
\underbrace{\gamma_4 \mathcal{P}^\dagger \gamma_4 \mathcal{P}}_{\mathcal{T}}u_a(x)
\big(
u_b(x)\,\Gamma^\prime\, d_c^T(x)
\big)
\big(
\bar{u}_{b'}(0)\,
\gamma_4 \Gamma^{\prime*} \gamma_4\,
\bar{d}_{c'}^T(0)
\big)\right]
\Big| 0 \Big\rangle \\
&=
a^3\sum_{\mathbf{x}}
e^{-i\mathbf{p}\cdot\mathbf{x}}\;
\epsilon_{abc}\,\epsilon_{a'b'c'}\;
\Big\langle 0 \Big|
\text{Tr}\left[\underbrace{\bar{u}_{a'}(0)\,\mathcal{T}u_a(x)}_{\text{free quarks}}
\big(
u_b(x)\,\Gamma^\prime\, d_c^T(x)
\big)
\big(
\bar{u}_{b'}(0)\,
\gamma_4 \Gamma^{\prime*} \gamma_4\,
\bar{d}_{c'}^T(0)
\big)\right]
\Big| 0 \Big\rangle \,.
\end{aligned}
\label{eq:C2pt}
\end{equation}

Kinematic enhancement can be applied to both the diquark and the free quark. Previously, the parity projection $\mathcal{P}_+\equiv(1+\gamma_t)/2$ was commonly used, as it has maximum overlap with the proton ground state at zero momentum. However, as pointed out in \cite{Zhang:2025hyo}, parity is no longer a good quantum number at non-zero momenta, particularly at large ones. 

More specifically, before projection, the nucleon two-point correlation function constructed with the interpolator $\hat{J}$ can be expressed as~\cite{Lee:1998cx,Stokes:2016oqk},
\begin{align}
    \langle0|\hat{J}^\dagger(t)\hat{J}|0\rangle =\sum_i |\lambda_{i,+}|^2\frac{\gamma\cdot P + M_i^+}{2E_i^+} e^{-E_i^+t} + \sum_i|\lambda_{i,-}|^2\frac{\gamma\cdot P - M_i^-}{2E_i^-} e^{-E_i^-t},
\end{align}
where $\lambda_{i,\pm}$ are the overlap of the interpolator with the $i$-th positive (negative) parity states, and $E_i^\pm$ and $M_i^\pm$ are the corresponding energies and masses. At rest,  the projectors $\mathcal{P}_\pm$ can eliminate the entire sector of negative (positive) parity states. But when boosted, this is not possible. On the other hand, for a general spin matrix $\mathcal{T}$,
\begin{align}
\mathcal{T}=b_0+\sum_{i=1}^4~c_i\gamma_i+\sum_{i=1}^4~d_{ij}\sigma_{ij}+e_0\gamma_5+\sum_{i=1}^4~f_i\gamma_5\gamma_i,
\end{align}
the projection of the two-point correlator is
\begin{align}
    {\rm Tr} [\mathcal{T}\langle0|\hat{J}^\dagger(t)\hat{J}|0\rangle]=&\sum_i |\lambda_{i,+}|^2\frac{c_3P_z+c_4E_i^+ + b_0M_i^+}{2E_i^+} e^{-E_i^+t} \nonumber\\
    &+ \sum_i |\lambda_{i,-}|^2\frac{c_3P_z+c_4E_i^- - b_0M_i^-}{2E_i^-} e^{-E_i^-t}.
\end{align}
Thus the requirement of a positive-definite spectral expansion results only in constraints for its coefficients
\begin{align}
    c_4\geq|b_0|, \qquad c_3\geq0,
\end{align}
while the coefficients $d_{ij}$ and $f_i$ are related to the polarization, which does not contribute to the spin-averaged two-point correlation functions. The $e_0$ factor will not contribute unless $\hat{J}^\dagger_{\rm snk}$ and $\hat{J}_{\rm src}$ operators are defined with opposite parities to each other, for example, when an extra $\gamma_5$ is inserted into only one of the diquarks.

This opens up other possibilities for $\mathcal{T}$ as well as the option to include linear combinations of multiple ``projectors'' at no additional computational cost like mixed parity, $\mathcal{T} = \mathcal{P}_+ - \mathcal{P}_- = \gamma_t$, which already provides kinematic enhancement. Another option which we shall explore in this paper is $\mathcal{T}=\gamma_+$, which is not a projector as $\gamma_+^2=0$, but also provides kinematic enhancement~\cite{LatticePartonCollaborationLPC:2025vhd}.

In the following, we considered different choices for both kernels, namely $\Gamma^\prime = C\gamma_5\Gamma$, where C is the charge conjugation matrix, with $\Gamma \in \{\mathbbm{1}, \gamma_t, \gamma_{+}\}$ and $\mathcal{T} \in \{\mathcal{P}_+, \gamma_t, \gamma_+\}$.

\begin{figure}[htbp]
    \centering
    \includegraphics[width=0.4\textwidth]{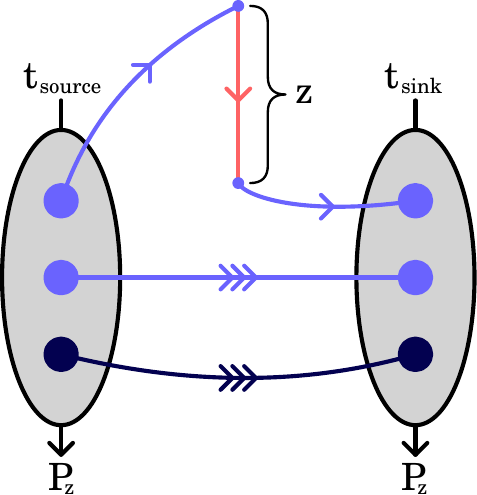}
    \caption{A nucleon 3pt contraction. Quark lines are coloured blue and dark blue for the up- and down quarks, respectively. The Wilson line is coloured red. Source and sink are placed at $t_\mathrm{source}$ and $t_\mathrm{sink}$, respectively. Both source and sink are boosted by $P_\mathrm{z}$, which is also the direction of the inserted Wilson link.}
    \label{fig:3pt_correlator}
\end{figure}

When using $\Gamma'=C\gamma_5\gamma_+$, the quarks forming the diquark are projected onto their ``$+$''-components. The free quark is projected likewise with $\mathcal{T}=\gamma_+$. This choice creates an additional factor of two improvement in variance compared to $(\mathcal{T},\Gamma)=(\gamma_t,\gamma_t)$ or $(\mathcal{T},\Gamma)=(\gamma_t,\gamma_z)$, as both information is used. In addition, this reduces the original degrees of freedom by projecting onto the ``$+$''-components only. This cancels all ``$-$''-components of the propagators in contractions, e.g. a nucleon three-point function, where the operator $O(x_{\text{ins}})=\bar{u}^{\delta'}_{d'}(x_{\text{ins}})K^{d'd}_{\delta'\delta}u^{\delta}_{d}(x_{\text{ins}})$ has been inserted.

\begin{align}
\label{eq:C3pt}
C^{u}_{3\text{pt},K}(\vec{p},\,t_{\text{ins}},\,t_{\text{sep}},\,z)
&= a^6 
\sum_{\vec{x}_{\text{snk}},\,\vec{x}_{\text{ins}}}
e^{-i\vec{p}\cdot(\vec{x}_{\text{snk}}-\vec{x}_{\text{src}})}
\,\epsilon_{abc}\,\epsilon_{a'b'c'}
(\mathcal{T})_{\alpha\alpha'}
(C\gamma_5\gamma_+)_{\beta\gamma}
(C\gamma_5\gamma_-)_{\beta'\gamma'}
\nonumber\\
\times
\Bigl\langle
u^{\alpha}_{a}(x_{\text{snk}})\,
&u^{\beta}_{b}(x_{\text{snk}})\,
d^{\gamma}_{c}(x_{\text{snk}})\,
\bar{u}^{\delta'}_{d'}(x_{\text{ins}})\,
K^{d'd}_{\delta'\delta}\,
u^{\delta}_{d}(x_{\text{ins}})\,
\bar{u}^{\beta'}_{b'}(x_{\text{src}})\,
\bar{d}^{\gamma'}_{c'}(x_{\text{src}})\,
\bar{u}^{\alpha'}_{a'}(x_{\text{src}})
\Bigr\rangle. 
\nonumber \\
&=
a^6 \sum_{\vec{x}_{\text{snk}},\,\vec{x}_{\text{ins}}}
e^{-i\vec{p}\cdot(\vec{x}_{\text{snk}}-\vec{x}_{\text{src}})}
\,\epsilon^{abc}\,\epsilon^{a'b'c'}
(\mathcal{T})_{\alpha\alpha'}
(C\gamma_5\gamma_+)_{\beta\gamma}
(C\gamma_5\gamma_-)_{\beta'\gamma'}
\nonumber\\
&\times
D^{cc'}_{\gamma\gamma'}(x_{\text{snk}},x_{\text{src}})
\nonumber\\
&\times
\Bigl[
  {U^{\alpha\delta'}_{ad'}(x_{\text{snk}},x_{\text{ins}})}\,
  K^{d'd}_{\delta'\delta}\,
  U^{\delta\alpha'}_{da'}(x_{\text{ins}},x_{\text{src}})\,
  U^{\beta\beta'}_{bb'}(x_{\text{snk}},x_{\text{src}})
\nonumber\\
&\quad
- {U^{\alpha\delta'}_{ad'}(x_{\text{snk}},x_{\text{ins}})}\,
  K^{d'd}_{\delta'\delta}\,
  U^{\delta\beta'}_{db'}(x_{\text{ins}},x_{\text{src}})\,
  U^{\beta\alpha'}_{ba'}(x_{\text{snk}},x_{\text{src}})
\nonumber\\
&\quad
+ {U^{\beta\delta'}_{bd'}(x_{\text{snk}},x_{\text{ins}})}\,
  K^{d'd}_{\delta'\delta}\,
  U^{\delta\beta'}_{db'}(x_{\text{ins}},x_{\text{src}})\,
  U^{\alpha\alpha'}_{aa'}(x_{\text{snk}},x_{\text{src}})
\nonumber\\
&\quad
- {U^{\beta\delta'}_{bd'}(x_{\text{snk}},x_{\text{ins}})}\,
  K^{d'd}_{\delta'\delta}\,
  U^{\delta\alpha'}_{da'}(x_{\text{ins}},x_{\text{src}})\,
  U^{\alpha\beta'}_{ab'}(x_{\text{snk}},x_{\text{src}})
\Bigr]\,.
\end{align}
When projecting all quarks in the source onto their ``$+$''-components via $(\mathcal{T},\Gamma)=(\gamma_+,\gamma_+)$, all ``$-$''-component contractions of the resulting propagators with the source kernels become exactly 0, only the ``$+$''-components contribute. This allows us to reduce the number of inversions to obtain the propagator connecting source and sink, in our case from a point source, by half. This reduction also applies to the propagator connecting the inserted operator to the sink, if it is obtained via the sequential source technique \cite{Martinelli:1988rr}, even if the sink interpolator does not project its quarks onto their ``$+$''-components. The propagators obtained from the ``$+$''-projected source interpolator ensure that in the contraction with the sink, only ``$+$''-components contribute, making all ``$-$''-components in the sequential source obsolete. This reduction in inversions also applies to polarized hadrons and partons, as polarization does not mix ``$+$'' and ``$-$'' components. Explicitly, ``$-$''-components obtained as $q_-=\tfrac{1}{\sqrt{2}}\gamma_+\gamma_-q$ still cancel after combining $\gamma_+$ with the helicity or transversity projector:
\begin{align}
    \gamma_+\gamma_5\gamma_+\gamma_- &= 0 \nonumber\\
    \gamma_+\gamma_5\gamma_\perp\gamma_+\gamma_- &=0
\end{align}
This applies neither at the source nor the sink if $(\mathcal{T},\Gamma)=(\gamma_t,\gamma_+)$ is chosen, as contractions involving the free quark mix ``$+$'' and ``$-$''-components. An example for this is shown in (\ref{eq:C3pt}), where the propagators connecting the operator with the sink that contract with the free quark kernel require all 12 indices. One such contraction, the first summand, is presented in Figure \ref{fig:3pt_correlator}, with the down quark propagator shown in dark blue, and the up quark ones in light blue. To fully exploit this additional improvement in computation time, we consider only $(\mathcal{T},\Gamma)=(\gamma_+,\gamma_+)$ for the case of ``$+$''-improvement. We summarize all tuples of kernels for our investigation in Table \ref{tab:tuple_choices}.

\begin{table}[h!]
\centering
\begin{tabular}{c c}
\hline
Kernels $(\mathcal{T},\Gamma)$ & Name \\
\hline
$(\mathcal{P}_+,\mathbbm{1})$ & Conventional \\
$(\gamma_+,\gamma_+)$ & $\gamma_+$ enhanced \\
$(\gamma_t,\gamma_t)$ & $\gamma_t$ enhanced \\
\hline
\end{tabular}
\caption{Choices of interpolator kernels $(\mathcal{T},\Gamma)$ used in this work.}
\label{tab:tuple_choices}
\end{table}

\subsection{Nucleon PDFs with LaMET}

The unpolarized nucleon quark isovector PDF is a well-studied quantity within the LaMET formalism, as described below. In this work, we do not repeat the entirety of the calculations involved, but restrict ourselves to computing only the quasi-PDF matrix elements as examples. 

\subsubsection{The unpolarized nucleon PDF}

In LaMET, the $\overline{\rm MS}$ lightcone PDF $f(x,\mu)$ is calculated from the following effective theory expansion~\cite{Ji:2013dva,Xiong:2013bka,Ji:2014gla,Izubuchi:2018srq,Ji:2020ect},
\begin{equation}
    f(x, \mu)=\int_{-\infty}^{\infty} \frac{d y}{|y|} \tilde{C}\left(\frac{x}{y}, \frac{\mu}{y P_z}\right) \tilde{f}\left(y, P_z, \mu\right)+\mathcal{O}\left(\frac{\Lambda_{\mathrm{QCD}}^2}{\left(x P_z\right)^2}, \frac{\Lambda_{\mathrm{QCD}}^2}{\left((1-x) P_z\right)^2}\right)\,,
    \label{eq:matching}
\end{equation}
where $\tilde{C}\left(\frac{x}{y}, \frac{\mu}{y P_z}\right)$ is a perturbative matching kernel and $\tilde{f}\left(y, P_z, \mu\right)$ is the quasi-PDF:
\begin{equation}
    \begin{aligned}
& \tilde{f}\left(y, P^z , \mu\right)=P^z \int \frac{d z}{2 \pi} e^{i z\left(y P^z\right)} \tilde{h}_R\left(z, P^z, \mu\right), \\
& \tilde{h}_R\left(z, P^z , \mu\right) = \frac{1}{2 E}\langle P | \bar{\psi}(z) \gamma^t \psi(0)|P\rangle\,.
\end{aligned}
\end{equation}
The quasi-PDF matrix elements $\tilde{h}\left(z, P^z , \mu\right)$ are calculated on the lattice and renormalized and matched to the $\overline{\rm MS}$ scheme. On the lattice,
\begin{align} \label{eq:quasiPDF}
\tilde{h}(z,P_z,1/a)&=\frac{1}{2 E}\langle P|\bar{\psi}(z) \gamma^{t}\mathcal{W}[z,0] \psi(0)|P \rangle, 
\end{align}
where $\mathcal{W}[z,0]$ is the straight Wilson line running from $0$ to $z$. Repeating this calculation with otherwise identical parameter sets but different interpolators allows us to assess their merits. While we perform these calculations only for this specific quantity, the results will have implications for all boosted hadrons. For example, precise results for large $P^z$ are crucial for a reliable infinite momentum extrapolation in LaMET. See Eq.~\eqref{eq:matching}.

\subsubsection{Renormalization}
\label{sec:renormalization}
The bare correlation function in Eq.~\eqref{eq:quasiPDF} suffers from linear divergences due to the Wilson self-energy and logarithmic UV divergences stemming from the UV regulating lattice spacing~\cite{Dotsenko:1979wb,Craigie:1980qs,Dorn:1986dt,Ji:2017oey,Ishikawa:2017faj,Green:2017xeu}. Both need to be removed correctly for a proper renormalization. We choose the hybrid scheme~\cite{Ji:2020brr} which renormalizes short distances $z$ in a ratio scheme \cite{Radyushkin:2017cyf, Orginos:2017kos, Radyushkin:2018cvn} and long ones via self-renormalization \cite{LatticePartonLPC:2021gpi}:
\begin{align}\label{eq:hybridscheme}
\tilde{h}_R(z,P_z, \mu)
=&\frac{\tilde{h}(z,P_z,1/a)}{\tilde{h}(z,P_z=0,1/a)}\theta(z_s-|z|)+	\eta_s\frac{\tilde{h}(z,P_z,1/a)}{Z_R(z,1/a)} \theta(|z|-z_s)\,.
\end{align}
The short-distance scale $z_s$ is introduced as an artificial separator between short and long distances, {which will be canceled by the $z_s$ dependence in the corresponding perturbative matching kernel}. It can be varied in the analysis to reduce bias associated with its choice. $\eta_s$ is a normalization factor that guarantees the continuity of renormalization at $z=z_s$ during the transition from perturbative to non-perturbative region.
At small $z\leq z_s$, we divide the correlation functions at large momenta by the same correlation functions in the rest frame. At large distances, the self-renormalization factor $Z_R(z,1/a)$ is used to remove UV divergences while preserving IR physics. The UV physics can be extracted from the short distance bare matrix elements in the rest frame $\tilde{h}(z,P_z=0,1/a)$  at multiple lattice spacings. Motivated by perturbation theory it is assumed to be of functional form 
\begin{align}\label{eq:lnbrMEFit}
    &\ln \tilde{h}(z,P_z=0,1/a) = \frac{kz}{a\ln(a\Lambda_{\rm{QCD}})}\!+\!g(z)\!+\!f(z)a^2\!\notag\\
    &+\!\frac{3C_F}{11-2N_f/3}\ln\left[\frac{\ln{[1/(a\Lambda_{\rm{QCD}})]}}{\ln{[\mu/\Lambda_{\rm{QCD}}]}}\right]\!+\!\ln\left[1\!+\!\frac{d}{\ln(a\Lambda_{\rm{QCD}})}\right]\,.
    \end{align}
The first term includes the linear divergence stemming from the Wilson line self-energy. The second term, $g(z)$, contains the non-perturbative information necessary for renormalization as well as a renormalon ambiguity term, which we will have to remove later. The third term, $f(z)a^2$, accounts for discretization errors. The last two terms result from the resummation of leading and sub-leading logarithms, affecting normalization at different lattice spacings. As higher order effects and further lattice artifacts cannot be removed entirely, $\Lambda_{\mathrm{QCD}}$ and $d$ are treated as fit parameters as well~\cite{LatticePartonLPC:2021gpi}. 
The normalized long distance renormalization coefficient $Z_R(z,1/a)/\eta_s$ is defined as 
{\begin{align}\label{eq:physcalfit}
\frac{Z_R(z,1/a)}{\eta_s} &= \tilde{h}(z_s,P_z=0,1/a)\exp\left[-{\frac{k(|z|-z_s)}{a\ln(a\Lambda_{\rm QCD})}-m_0(|z|-z_s)}\right]\,,
\end{align}}
where the renormalon mass $m_0$ is extracted via a linear fit to the difference of the continuum perturbative $\overline{\mathrm{MS}}$ result $C_{0,\mathrm{NLO}}^{\overline{\mathrm{MS}}}(z)$ and $g(z)$ obtained from \eqref{eq:lnbrMEFit}
\begin{align}
    g(z)-\ln(C_{0,\mathrm{NLO}}^{\overline{\mathrm{MS}}}(z))=m_0 z\,
\end{align}
The fixed order (NLO) result was obtained in \cite{Izubuchi:2018srq} and is given as
\begin{align}
\label{eq:oneloopPZ0}
C_{0,\mathrm{NLO}}^{\overline{\mathrm{MS}}}(z)=1+
\frac{\alpha_s C_F}{2\pi}\left(\frac{3}{2}\ln\left(z^2\mu^2e^{2\gamma_E}/4\right)+\frac{5}{2}\right).
\end{align}
{Note that the determination of $m_0$ is not unique due to the renormalon ambiguity, and needs to be properly regularized when matched to the lightcone PDF to avoid introducing extra linear corrections in $1/P_z$~\cite{Zhang:2023bxs,Holligan:2023rex}. However, the scope of this work only focuses on the statistical uncertainties of quasi-PDF lattice data, not including the perturbative matching to the lightcone PDFs.  It makes no difference if we simply use the fixed-order NLO result from \eqref{eq:oneloopPZ0}, as the ambiguity does not affect the relative statistical errors of the lattice data.}

\section{Lattice setup and techniques}
\label{section:setup}
To obtain \eqref{eq:quasiPDF} for the isovector case on the lattice, we compute the two-point function $C_{2\mathrm{pt}}(\vec{p}, t)$:
\begin{equation}
    C_{2\mathrm{pt}}(\vec{p},t) = \sum_{\vec{x}} e^{-i\vec{p}\cdot\vec{x}} \langle 0 | \mathcal{J}(\vec{x}, t)\, \overline{\mathcal{J}}(\vec{0}, 0) | 0 \rangle
\end{equation}
and the three-point function $C^{u-d}_{3\mathrm{pt}, V}(z, \vec{p},\tau,t)$, of which one contraction is illustrated in Figure \ref{fig:3pt_correlator},
\begin{equation}
    C^{u-d}_{3\mathrm{pt}}(z, \vec{p}, \tau, t_{\mathrm{sep}})  = \sum_{\vec{x}, \vec{y}} e^{-i\vec{p} \cdot (\vec{x}-\vec{y})} \langle 0 | \mathcal{J}(\vec{x}; t_{\rm sep})\, \mathcal{O}(\vec{y}, z;\tau)\, \overline{\mathcal{J}}(\vec{0}; 0) | 0 \rangle \,,
\end{equation}
where
\begin{equation}
    \mathcal{O}(\vec{y},z;\tau) = \bar{u}(\vec{y};\tau)\gamma^t W(\vec{y},\vec{y}+z\hat{z};\tau) u(\vec{y}+z\hat{z};\tau) 
- \bar{d}(\vec{y};\tau)\gamma^t W(\vec{y},\vec{y}+z\hat{z};\tau) d(\vec{y}+z\hat{z};\tau)\,,
\end{equation} 
containing a straight Wilson line $W$ of length $z$ in $\hat{z}$ (momentum) direction. The operator $ \mathcal{O}(\vec{y},z;\tau)$ is inserted at all insertion times $0<\tau<t_\mathrm{sep}$, where $t_{\mathrm{sep}}= t_{\mathrm{snk}} - t_{\mathrm{src}}$ is the source-sink separation. The interpolators $\mathcal{J}, \overline{\mathcal{J}}$ have been defined in Eq.~(\ref{eq:Interpolators}). \\
Inserting a complete set of states, the two-point function reads
\begin{equation}
    C_{2\mathrm{pt}}(\vec{p}, t_{\mathrm{sep}}) 
    = \sum_{n} \frac{|A_n|^2}{2E_n} e^{-E_n t_{\mathrm{sep}}},
\end{equation}
and the three-point function
\begin{equation}
    C^{u-d}_{3\mathrm{pt}}(z, \vec{p}, \tau, t_{\mathrm{sep}}) 
    = \sum_{m,n} \frac{A_m^* A_n}{4{E_m E_n}}
      \langle m | \mathcal{O}(z) | n \rangle \,
      e^{-E_m(t_{\mathrm{sep}}-\tau)} e^{-E_n \tau}\,.
\end{equation}
We then form their ratio
\begin{equation}
    R^{u-d}_{3\mathrm{pt},V}(z, \vec{p}, \tau, t_{\mathrm{sep}}) 
    = \frac{C^{u-d}_{3\mathrm{pt}}(z, \vec{p}, \tau, t_{\mathrm{sep}})}
           {C_{2\mathrm{pt}}(\vec{p}, t_{\mathrm{sep}})}
\end{equation}
and perform a two-state fit to both the two-point function and $R^{u-d}_{3pt,V}$ to extract the ground state matrix element
\begin{align}
C_{2\mathrm{pt}}(P_z, t_{\mathrm{sep}}) 
    &= c_0 e^{-E_0 t_{\mathrm{sep}}}+c_1 e^{-E_1 t_{\mathrm{sep}}}\\ \nonumber
    R^{u-d}_{3\mathrm{pt},V}(z, P_z, \tau, t_{\mathrm{sep}}) &= 
\tfrac{
\tilde{h}(z,P_z,1/a)
+ 2c_{01}(z)\cosh\!\left(\Delta E\left(\tau - \frac{t_{\mathrm{sep}}}{2}\right)\right)
e^{-\frac{\Delta E}{2} t_{\mathrm{sep}}}
+ c_{11}(z)\, e^{-\Delta E\, t_{\mathrm{sep}}}
}{
1 + \frac{c_1}{c_0} e^{-\Delta E\, t_{\mathrm{sep}}}
}\,,
\end{align}
where we have chosen $\vec{p}=(0,0,P_z)$. $\Delta E = E_1 - E_0$ is the energy gap between ground and first excited hadron state, $c_{ij}, c_k$ are fit parameters derived from the overlaps $c_{ij}\propto\bra{i}\mathcal{O}(z)\ket{j}$, $c_{k}\propto\bra{0}\mathcal{J}\ket{k}$ and $\tilde{h}(z,P_z,1/a)$ is the ground state matrix element we want to obtain.

We use three ensembles with open boundary conditions in time generated by the CLS collaboration, namely H102, S400 and N203 \cite{Bruno:2014jqa}. These use  $n_f=2+1$ $\mathcal{O}(a)$ non-perturbatively improved Sheikholeslami-Wohlert fermions and a tree-level Lüscher-Weisz gauge action.
Details about the ensembles can be found in Table~\ref{tab:3pt_params}, taken from \cite{RQCD:2022xux}.
We perform calculations for two different momenta. For renormalization, it is necessary to extract the matrix elements at nucleon momentum $P_z = 0\,\mathrm{GeV}$. In this case, the conventional interpolator is the superior choice. We, therefore, calculate the matrix element in the rest frame using it exclusively.
The interpolator comparison is made at a nucleon momentum of $P_z\approx2.5\,\mathrm{GeV}$ across the three lattice ensembles, for details see Table~\ref{tab:3pt_params}. We repeat this calculation, using otherwise identical parameters, for three different combinations of kernels as specified in Table~\ref{tab:tuple_choices}.
For N203, we work with a single source per configuration, while for H102 and S400, we use two sources per configuration to increase statistics.

In order to improve the signal, two smearing techniques are applied.
First, we prepare two APE-smeared gauge fields with $\alpha_\mathrm{APE} = 0.615384$, one for the calculation of the Wilson line and one for quark smearing \cite{Albanese:1987}.
For the calculation of the Wilson line, we only apply two steps of APE smearing to preserve the UV fluctuations of the gauge field. This is crucial for the correct long range renormalization. To obtain the quark propagators, we apply source-sink momentum smearing \cite{Bali:2016lva} ($k_\mathrm{z} = -0.45$, $\alpha_\mathrm{z} = 0.25$, $n_\mathrm{steps} = \{250, 310, 445\}$ for H102, S400 and N203 respectively) where we use 25 APE smearing steps. Here, the APE smeared gauge field is used only to maintain covariance of the source/propagator and not for its physical content. Thus, we can afford more smearing steps to reduce noise stemming from gauge field fluctuations. We use point-to-all propagators to evaluate the two-point function and the sequential source technique for the three-point function \cite{Martinelli:1988rr}. One contraction contributing to the three-point function is illustrated in Figure \ref{fig:3pt_correlator}. Details on the chosen separation times, Wilson line lengths and numbers of measurements for the different momenta and ensembles can be found in Table \ref{tab:3pt_params}.

Most lattice operations (IO, smearing, inversion) are performed using the \texttt{QUDA} library \cite{Clark:2010, Babich:2011}.
In order to use \texttt{QUDA} in our scripts we made use of the \texttt{QUDA}-Python bindings package \texttt{PyQUDA} \cite{jiang:2024}.

\begin{table}[h]
    \begin{center}
    \begin{tabular}{ccccccccc}
    \hline
    \hline
    id & $\beta$ & $a[\mathrm{fm}]$  & $L^3 \times T$ & $\kappa_{l}$ & $\kappa_{s}$ & $m_{\pi,K}[\mathrm{MeV}]$ & $m_\pi L$ & $N_{used}$ \\
    \hline
    H102b & $3.4$  & $0.085$ & $32^3 \times 96$  & $0.136865$ & $0.136549339$ & $354$, $442$ & $4.89$ & $400$ \\
    S400 & $3.46$ & $0.075$ & $32^3 \times 128$ & $0.136964$ & $0.136702387$ & $354$, $445$ & $4.33$ & $400$ \\
    N203  & $3.55$ & $0.064$ & $48^3 \times 128$ & $0.137080$ & $0.136840284$ & $348$, $445$ & $5.39$ & $200$ \\
    \hline
    \hline
    \end{tabular}
    
   \end{center}
    \begin{center}
    \begin{tabular}{cccc}
    \hline
    \hline
    Ensemble & $t_\mathrm{sep}/a$ & $z_\mathrm{max}/a$ & $P_z\,[\mathrm{GeV}]\,:\,N_\mathrm{meas}$ \\
    \hline
    H102     & \{6, 8, 10\}       & 20                 & \{0: 400, 2.7: 800\}                  \\
    S400     & \{7, 9, 11\}       & 20                 & \{0: 400, 2.5: 800\}                  \\
    N203     & \{8, 11, 13\}      & 20                 & \{0: 200, 2.4: 200\} \\
    \hline
    \hline
    \end{tabular}
    
    \end{center}
    \caption{Details of the CLS ensembles used for the investigation of the lattice spacing dependence.\label{tab:3pt_params}}
\end{table}

\section{Results}
\label{section:results}
In this section, we present our numerical results obtained for nucleon observables using different interpolating operators, focusing on the impact of kinematic enhancements. Our analysis is structured into three subsections. We begin by examining excited-state contamination (ESC), a good understanding of which is crucial for a reliable determination of ground-state matrix elements. In \cite{Zhang:2025hyo}, the ESC was suspected to become more pronounced when using kinematic enhancement. We start by determining their size for the nucleon. Next, we investigate the possible choices for free quark and diquark projectors/kernels and their interactions. By varying the free quark projector, we assess how the Dirac structure of the interpolators affects the signal. Finally, we present a direct comparison of results obtained with different nucleon interpolators, both kinematically enhanced and conventional. This comparison allows us to quantify the potential gain in precision, i.e., the signal to noise ratio (SNR).

\subsection{Excited state contamination}
For a realistic comparison of data obtained for different nucleon interpolators, we perform a two-state fit for the $C_{\rm 3pt}/C_{\rm 2pt}$ ratio data. 
In these fits, contributions beyond the ground state are largely absorbed into the excited-state parameters, reducing their impact on the extracted ground-state matrix element.
The results are normalized by the matrix elements of local quark bilinears obtained in Ref.~\cite{RQCD:2022xux}. 
\begin{figure}[ht]
\centering

\begin{subfigure}{0.48\textwidth}
    \centering
    \includegraphics[width=\linewidth]{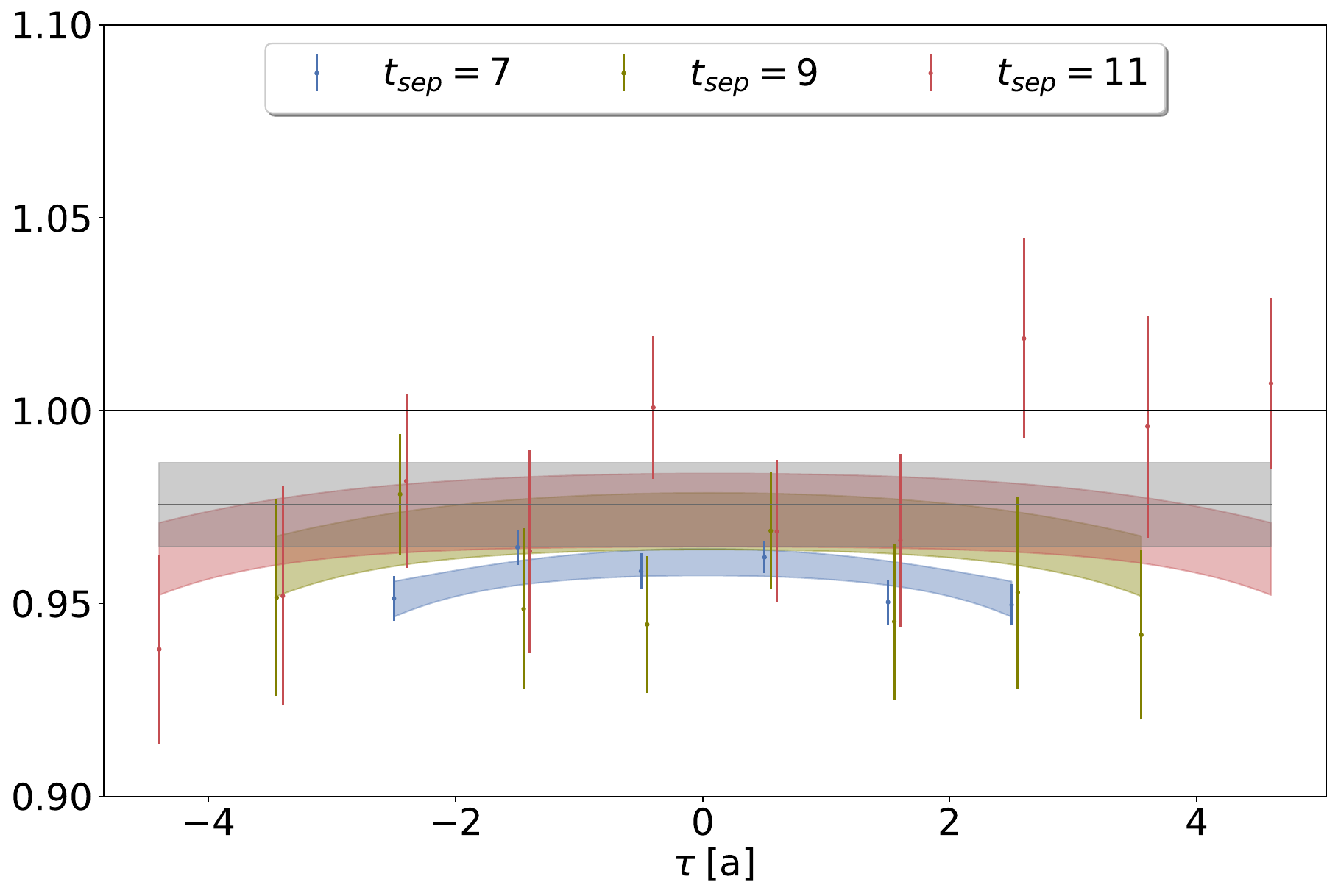}
    \caption{$z=0$, $\mathcal{T}=P_+$, $\Gamma=\mathbbm{1}$}
\end{subfigure}
\hfill
\begin{subfigure}{0.48\textwidth}
    \centering
    \includegraphics[width=\linewidth]{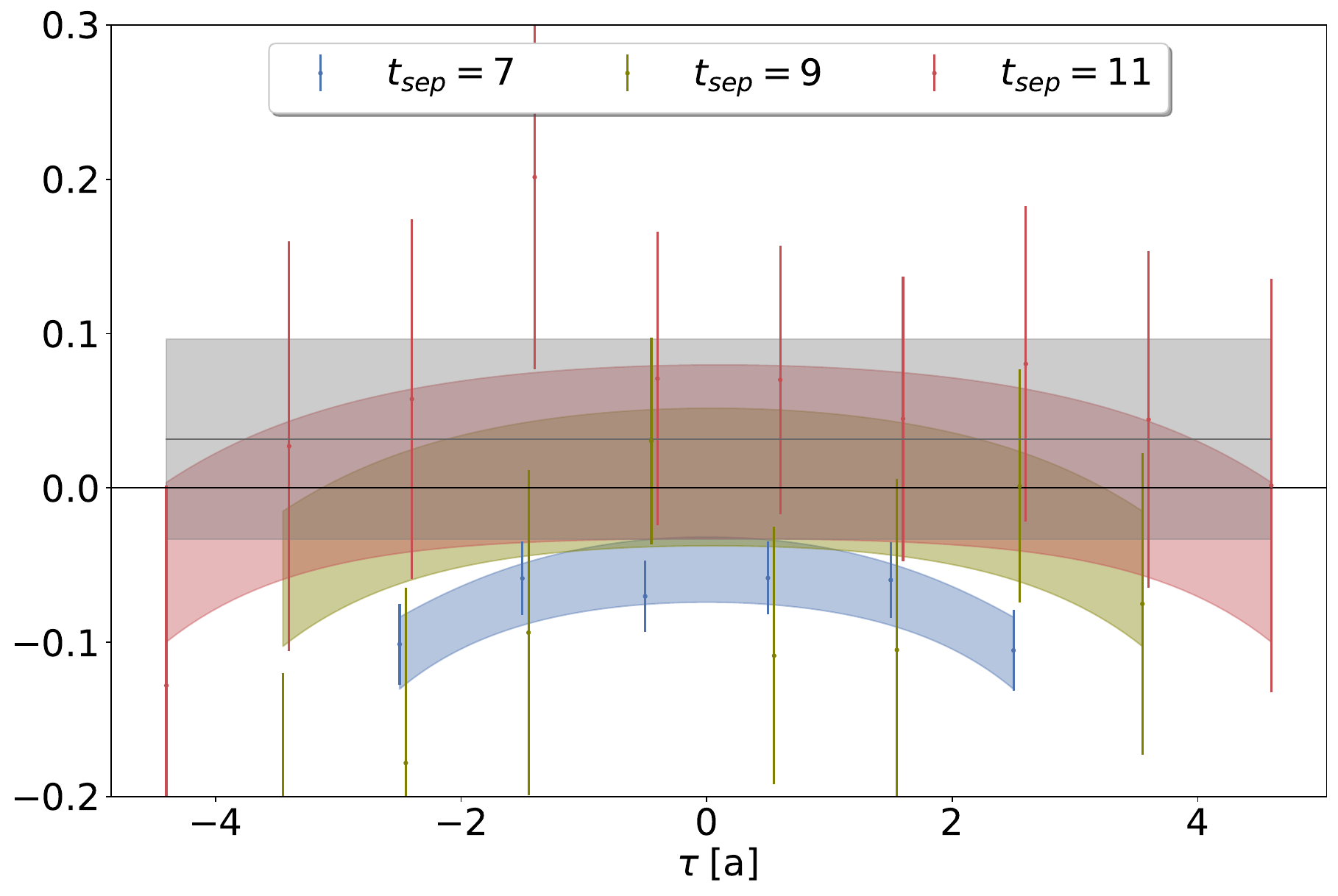}
    \caption{$z=8a$, $\mathcal{T}=P_+$, $\Gamma=\mathbbm{1}$}
\end{subfigure}

\vspace{0.5cm}

\begin{subfigure}{0.48\textwidth}
    \centering
    \includegraphics[width=\linewidth]{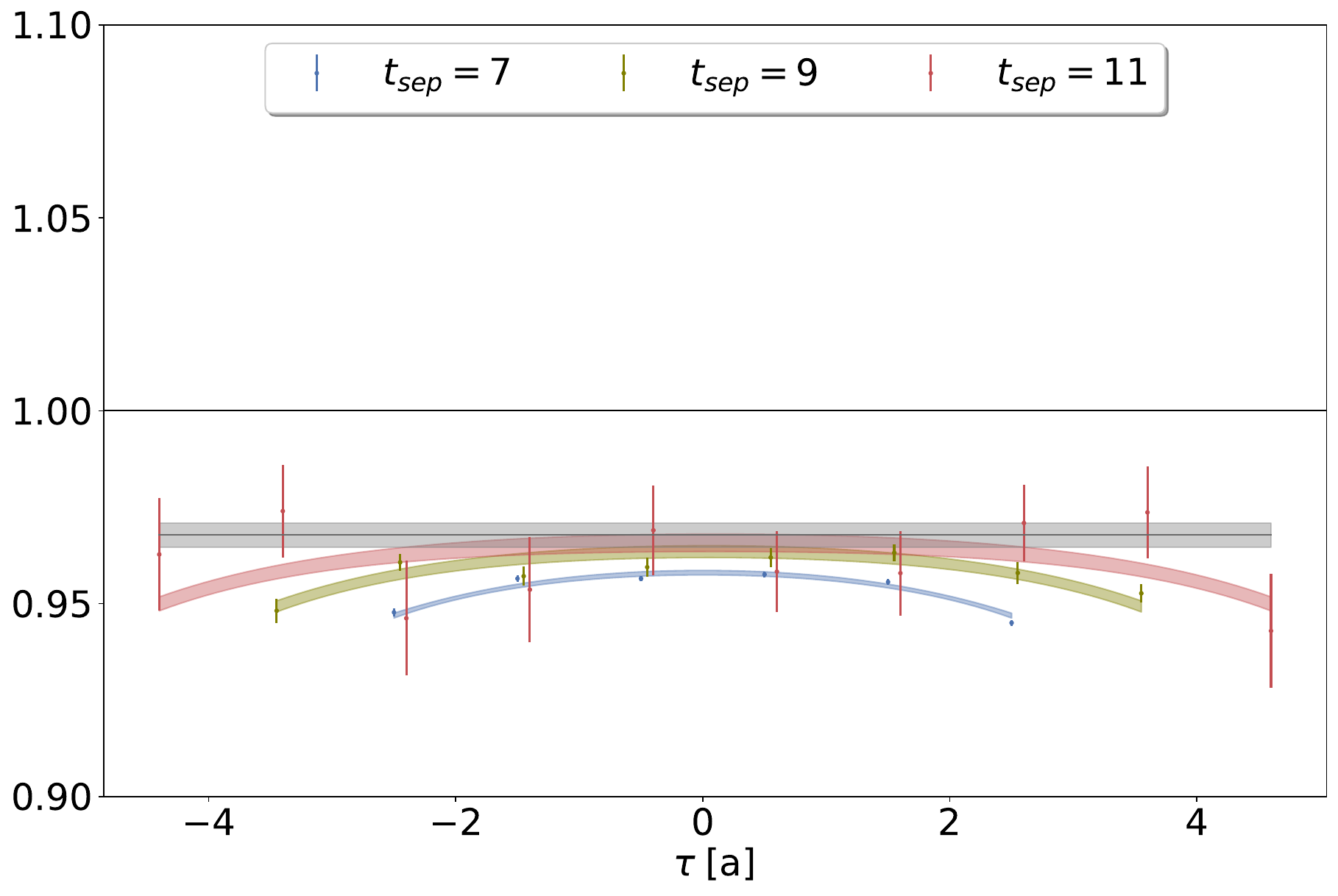}
    \caption{$z=0$, $\mathcal{T}=\gamma_t$, $\Gamma=\gamma_t$}
\end{subfigure}
\hfill
\begin{subfigure}{0.48\textwidth}
    \centering
    \includegraphics[width=\linewidth]{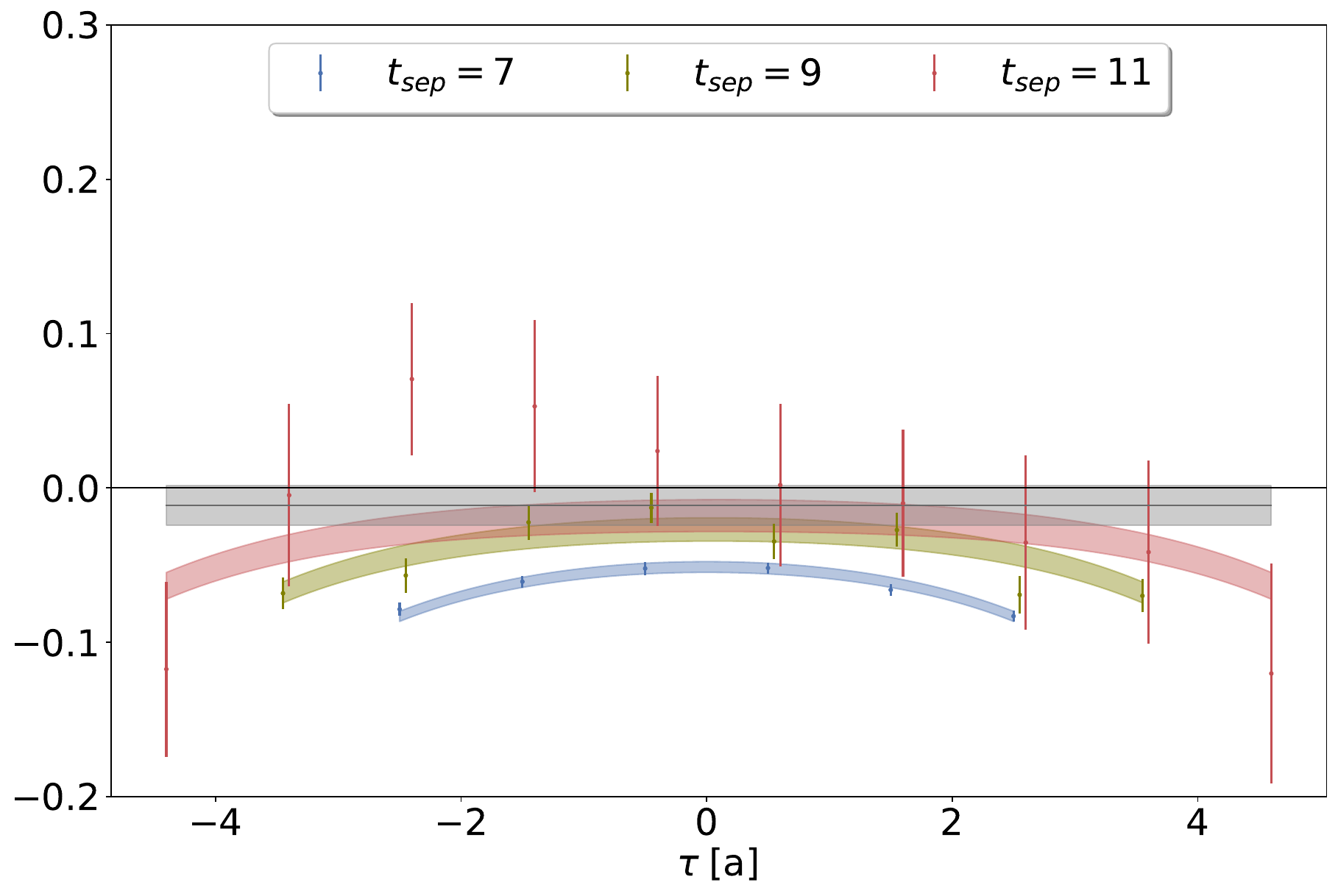}
    \caption{$z=8a$, $\mathcal{T}=\gamma_t$, $\Gamma=\gamma_t$}
\end{subfigure}

\vspace{0.5cm}

\begin{subfigure}{0.48\textwidth}
    \centering
    \includegraphics[width=\linewidth]{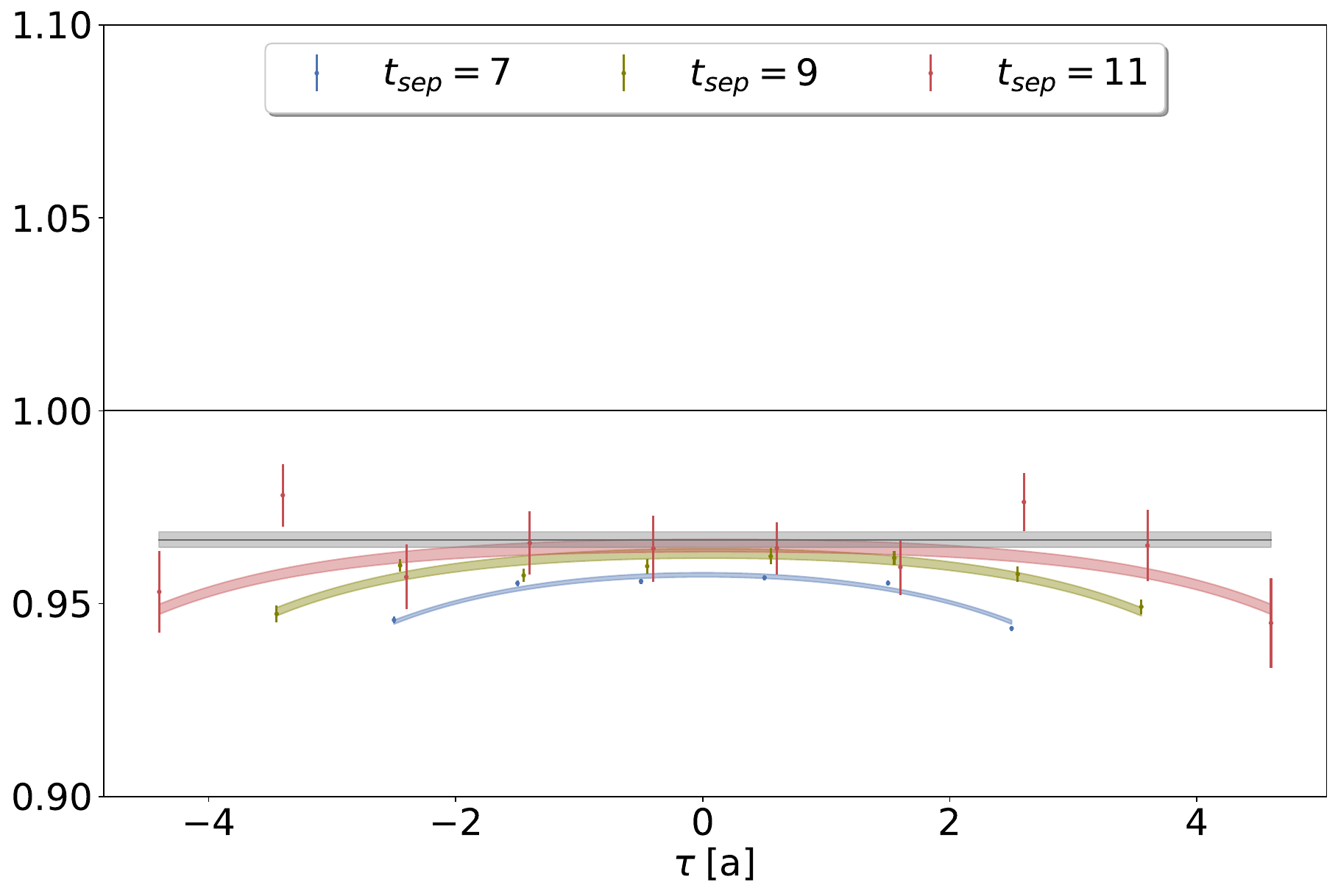}
    \caption{$z=0$, $\mathcal{T}=\gamma_+$, $\Gamma=\gamma_+$}
\end{subfigure}
\hfill
\begin{subfigure}{0.48\textwidth}
    \centering
    \includegraphics[width=\linewidth]{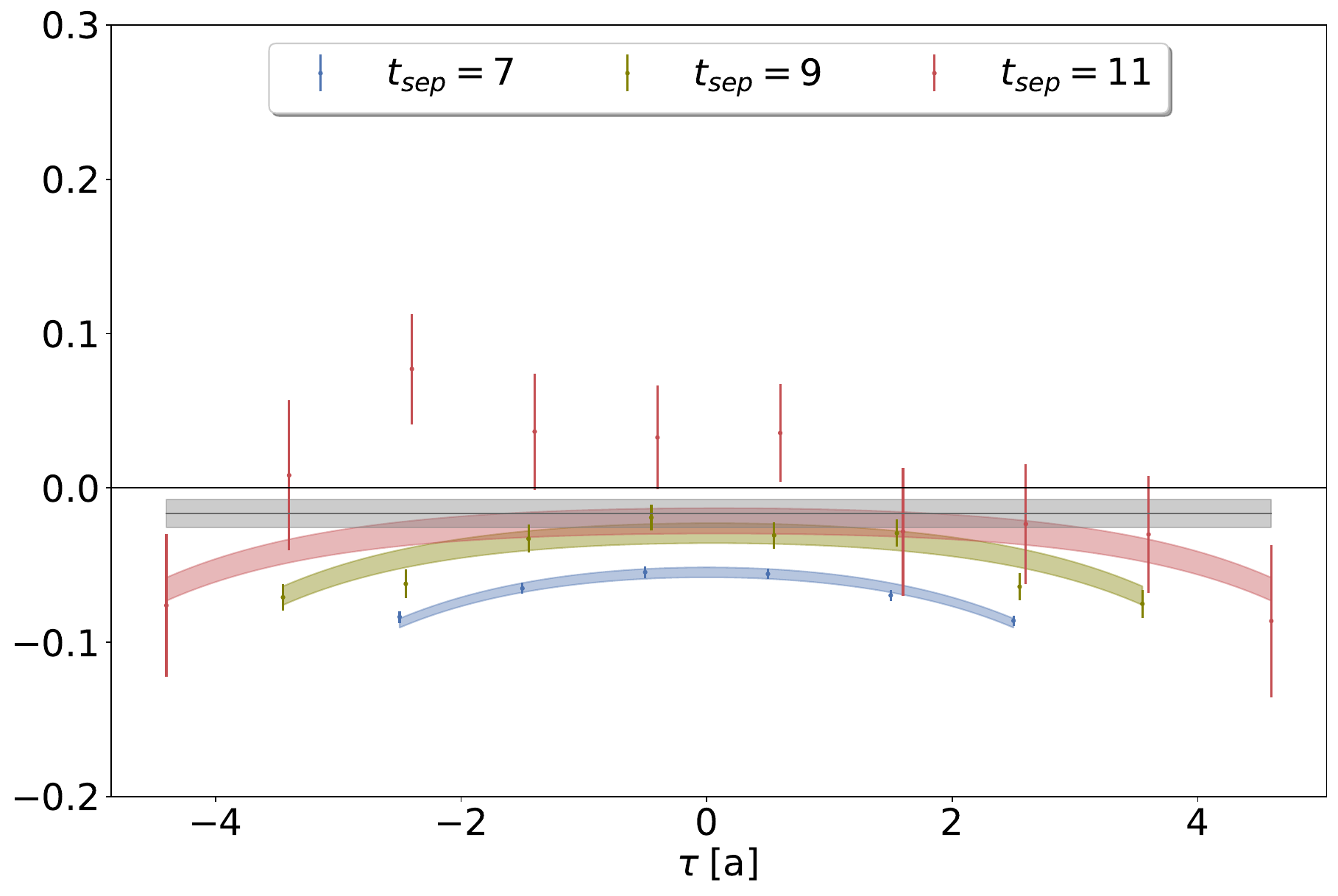}
    \caption{$z=8a$, $\mathcal{T}=\gamma_+$, $\Gamma=\gamma_+$}
\end{subfigure}

\caption{Two-state fit results (gray bands) for the isovector local ($z=0$, left column) and nonlocal ($z=8a$, right column) current at different separation times, locally normalized. Different interpolators are presented in rows, from top to bottom: conventional interpolator, $\gamma_t$ enhanced, $\gamma_+$ enhanced. The data was obtained from 400 measurements on the S400 ensemble at $P_z=2.5\,\mathrm{GeV}$.}
\label{fig:S400_multistate_3x2}
\end{figure}

\begin{figure}[htbp]
    \centering
    \includegraphics[width=0.9\textwidth, trim = 0 0 0 40, clip]{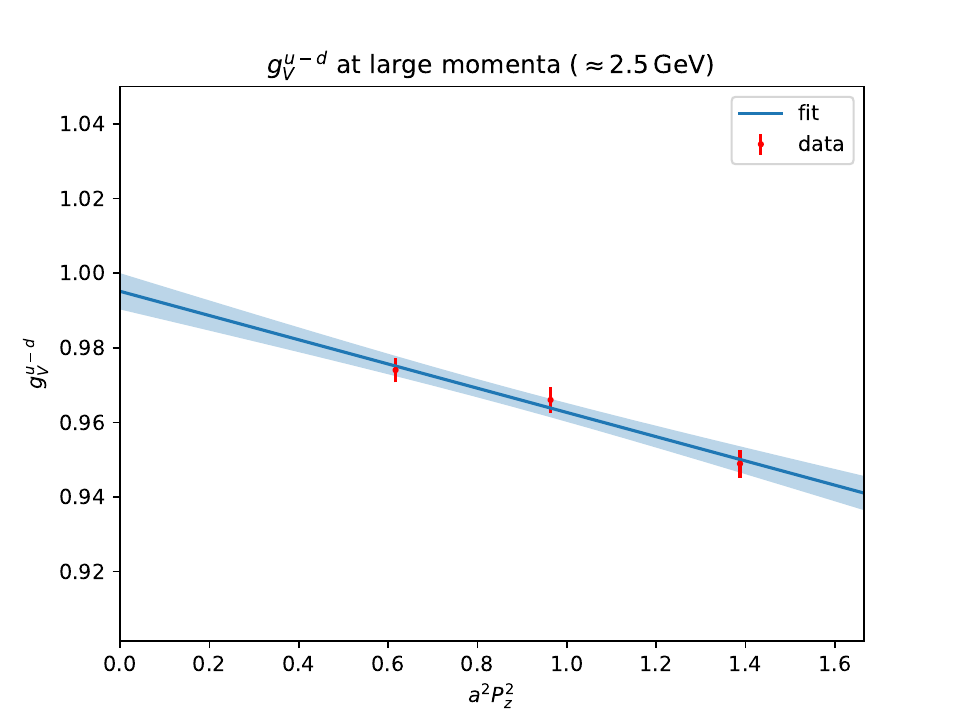}
    \caption{Renormalized local vector charge $g_V^{u-d}$, extrapolated linearly in $(aP_z)^2$.}
    \label{fig:discretization}
\end{figure}

In Figure \ref{fig:S400_multistate_3x2}, we compare the two-state fit results of different free quark and diquark kernel choices (top to bottom) for both the unpolarized nucleon quark isovector local current (left column) and a non-local current with Wilson line length $z=8a$ (right column). 
Within uncertainties we do not observe any difference in curvature
between the first row (standard interpolator) and the other two rows
(kinematically enhanced interpolators) implying that there is no
significant increase of ESCs when moving from the standard to the
kinematically enhanced interpolators. However, we do observe a
considerably improved SNR for the kinematically enhanced interpolators,
which we will discuss in detail in section \ref{subsec:Interpolator_comp}.
These results also allow to address the dependence on $aP^z$, i.e. the
size of discretization effects. These are expected to be of order
${\cal O}\big((aP_z)^2\big)$ when using ensembles with
${\cal O}(a)$ improved action. Figure \ref{fig:discretization} confirms this expectation,
indicating that still higher order discretization effects are small.

\newpage
\subsection{Interpolator comparison}

In this section, we compare our results obtained using different interpolators. We will first present the two-point and three-point functions used in the $C_{3pt}/C_{2pt}$ ratio separately, before turning to the latter. Finally, we present our results for renormalized matrix elements.
\label{subsec:Interpolator_comp}

\begin{figure}[htbp]
    \centering

    \begin{subfigure}[b]{0.48\textwidth}
        \centering
        \includegraphics[width=\textwidth]{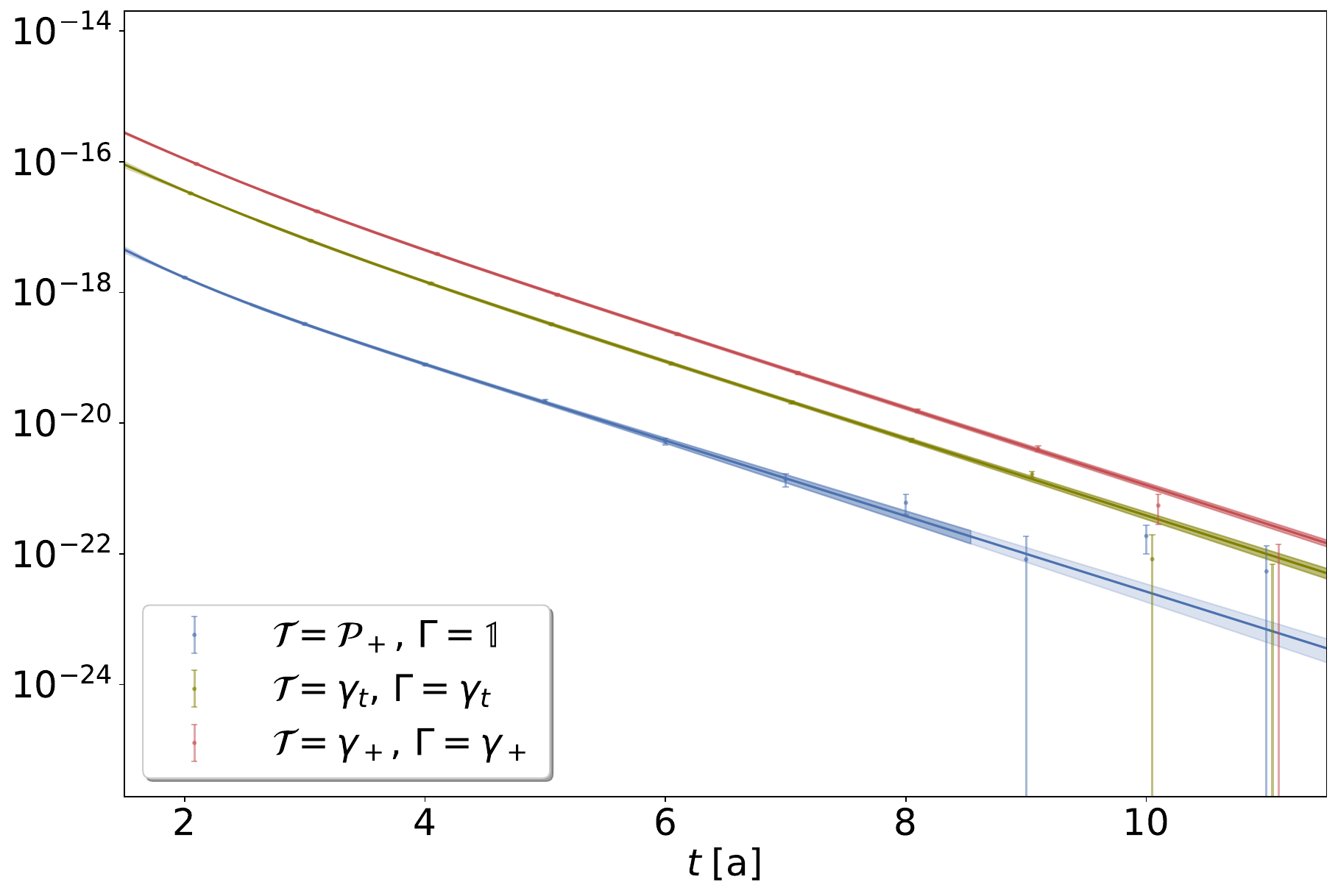}
        \caption{H102, $C_{2pt}$}
        \label{fig:2pt_H102}
    \end{subfigure}
    \hfill
    \begin{subfigure}[b]{0.48\textwidth}
        \centering
        \includegraphics[width=\textwidth]{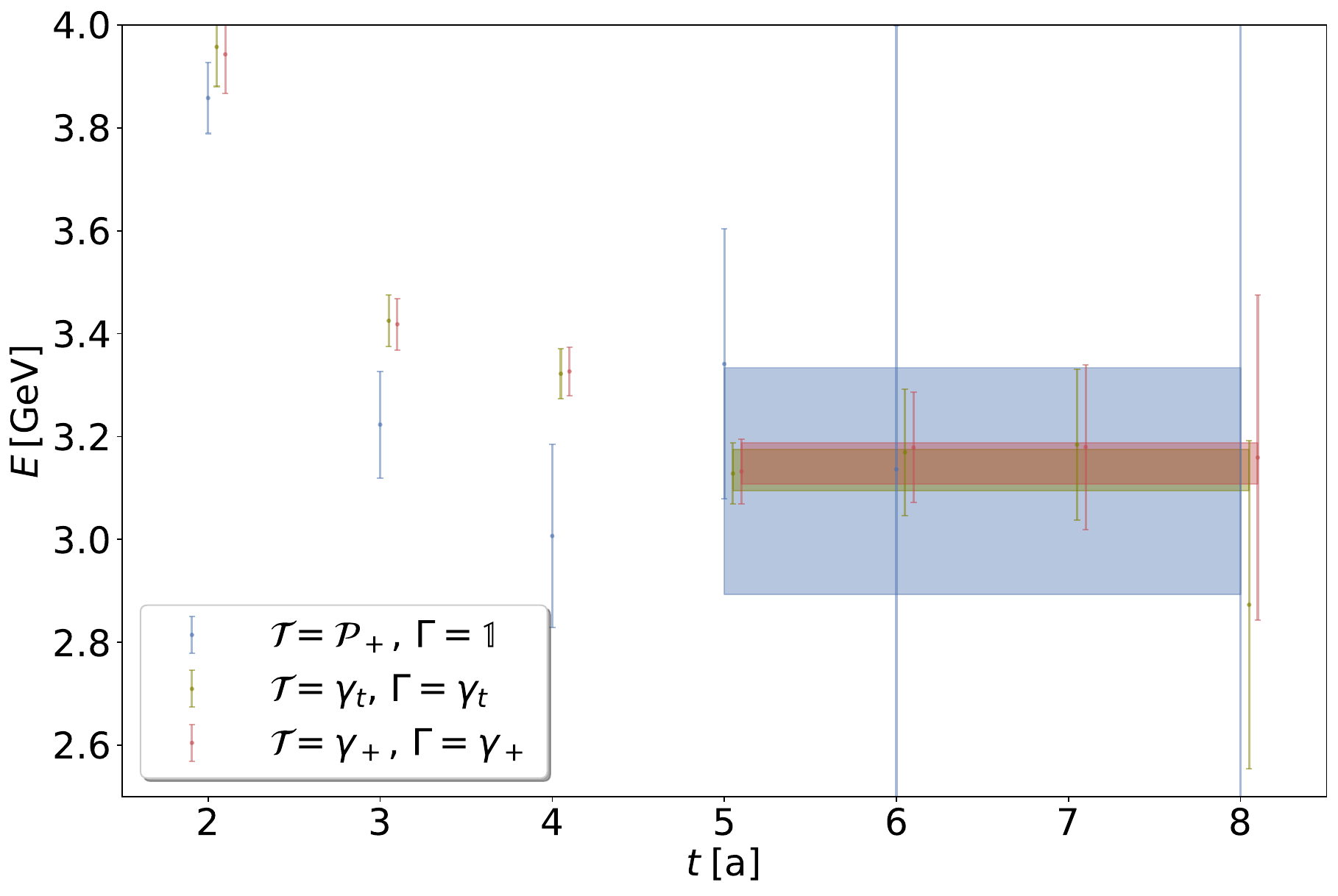}
        \caption{H102, $E_{\mathrm{eff}}$}
        \label{fig:Eeff_H102}
    \end{subfigure}

    \vspace{0.5cm}

    \begin{subfigure}[b]{0.48\textwidth}
        \centering
        \includegraphics[width=\textwidth]{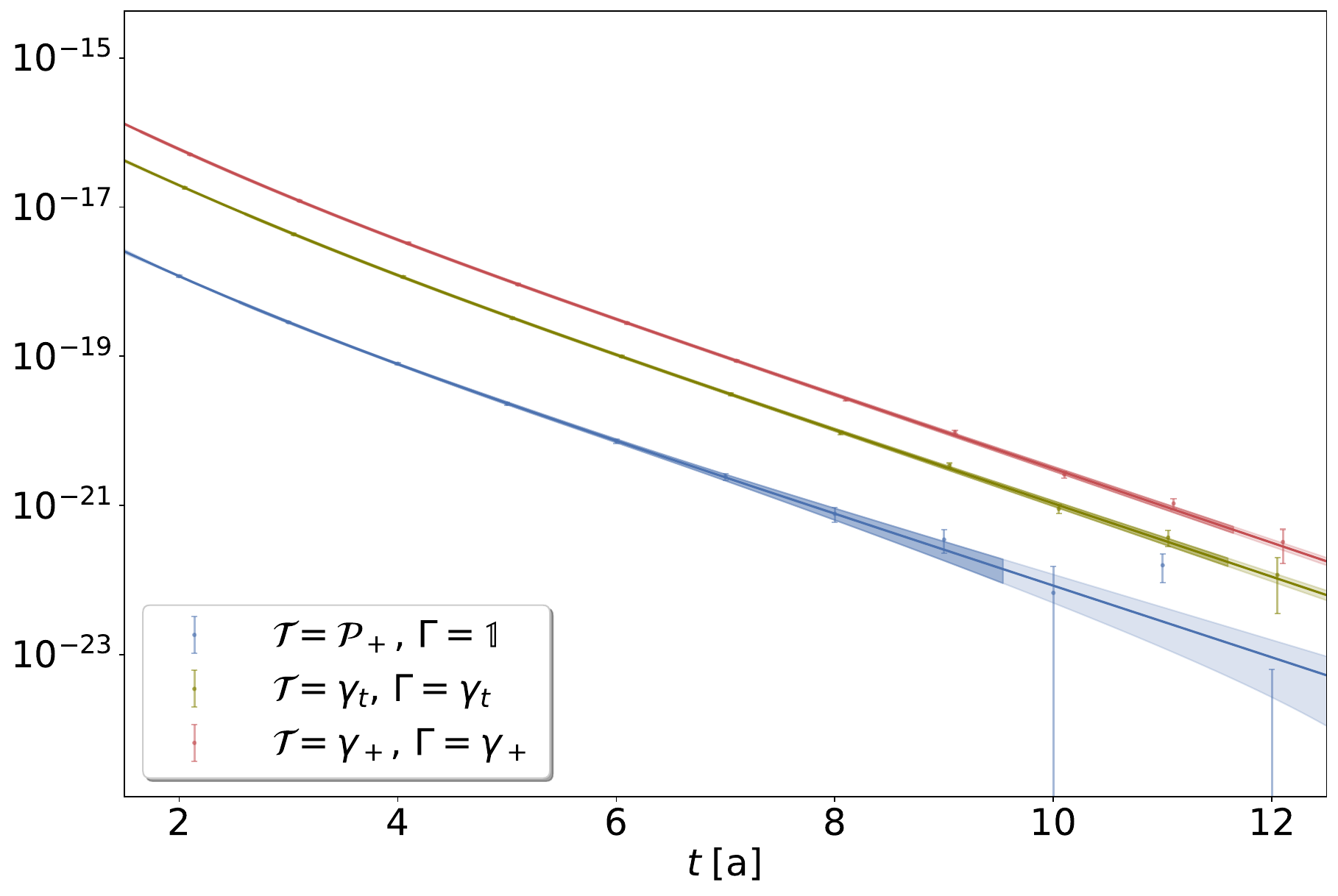}
        \caption{S400, $C_{2pt}$}
        \label{fig:2pt_S400}
    \end{subfigure}
    \hfill
    \begin{subfigure}[b]{0.48\textwidth}
        \centering
        \includegraphics[width=\textwidth]{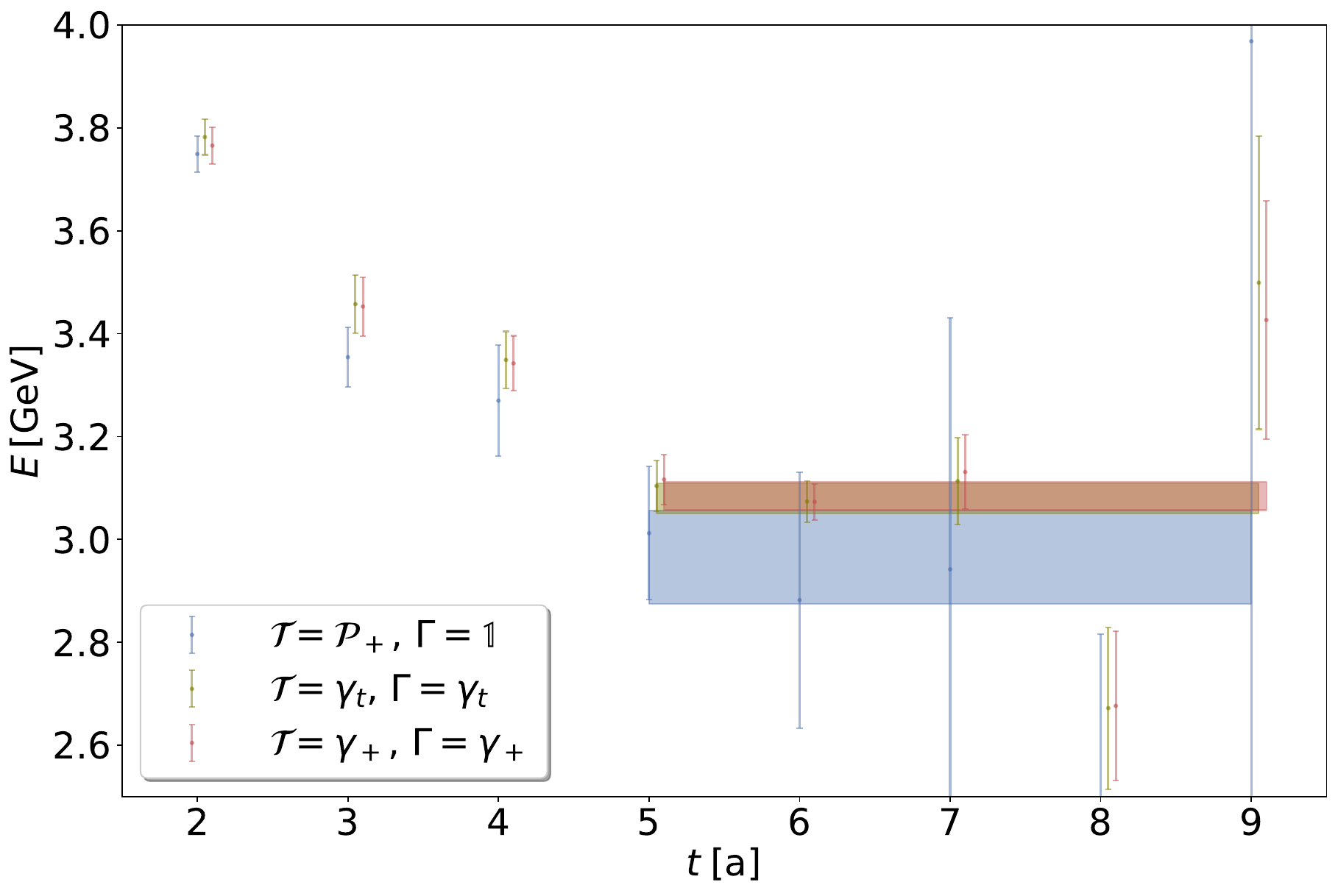}
        \caption{S400, $E_{\mathrm{eff}}$}
        \label{fig:Eeff_S400}
    \end{subfigure}

    \vspace{0.5cm}

    \begin{subfigure}[b]{0.48\textwidth}
        \centering
        \includegraphics[width=\textwidth]{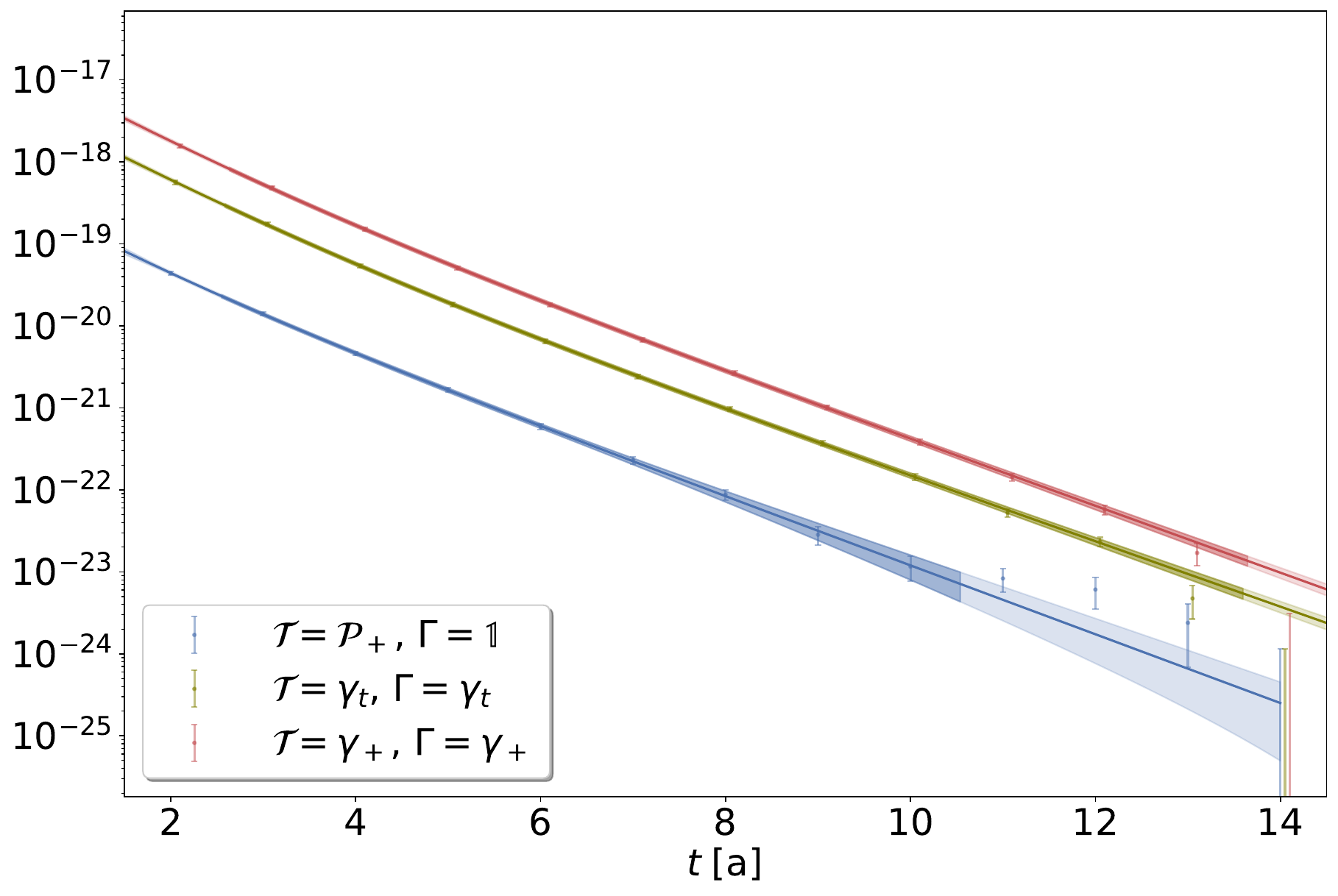}
        \caption{N203, $C_{2pt}$}
        \label{fig:2pt_N203}
    \end{subfigure}
    \hfill
    \begin{subfigure}[b]{0.48\textwidth}
        \centering
        \includegraphics[width=\textwidth]{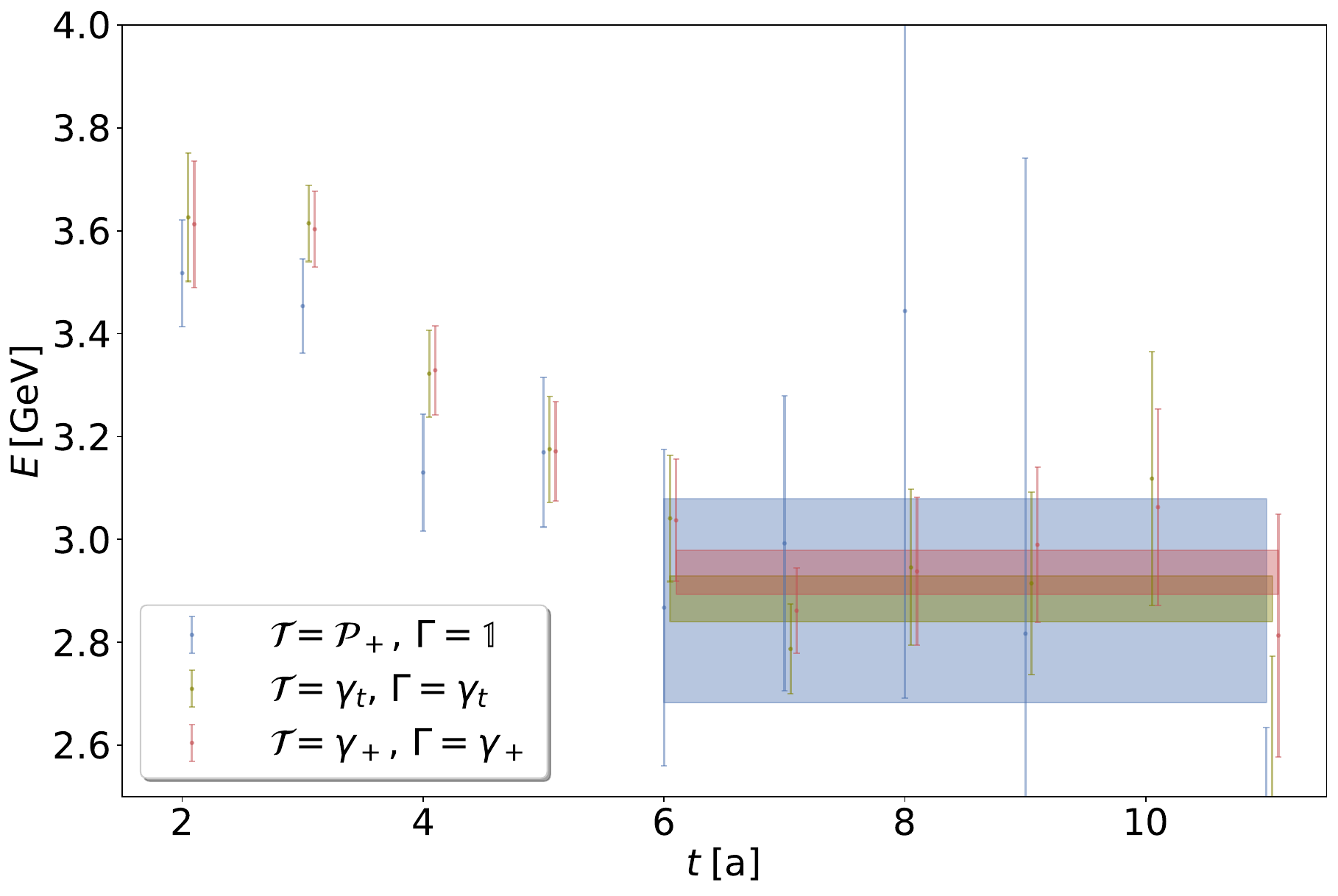}
        \caption{N203, $E_{\mathrm{eff}}$}
        \label{fig:Eeff_N203}
    \end{subfigure}

    \caption{
    Two-state fit results to the 2pt correlation functions (left column) and the corresponding effective energy plateaus (right column) for the H102 (top row), S400 (middle row) and N203 (bottom row) ensembles. The hadron momenta are close to $P_z=2.5$~GeV.}
    \label{fig:2pt_all}
\end{figure}

The two-point functions for the ensembles studied in this work are shown in Figure \ref{fig:2pt_all}. The results obtained using kinematically enhanced interpolators show a stronger signal and increased SNR, providing high precision results for larger times than the conventional interpolator. This is crucial for ground state matrix element extraction. At identical statistics (number of measurements), the maximum usable sink time $t$ is increased by $\approx50\%$. The SNR is improved considerably at smaller $t$, where both the signal from the conventional and the kinematically enhanced interpolators is usable, as found already in \cite{Zhang:2025hyo}. Both these facts contribute to an improved $C_{3pt}/C_{2pt}$ signal even before considering the potential enhancement to the three-point function.

\begin{figure}[ht]
\centering
\begin{subfigure}{0.78\textwidth}
    \centering
    \includegraphics[width=\linewidth]{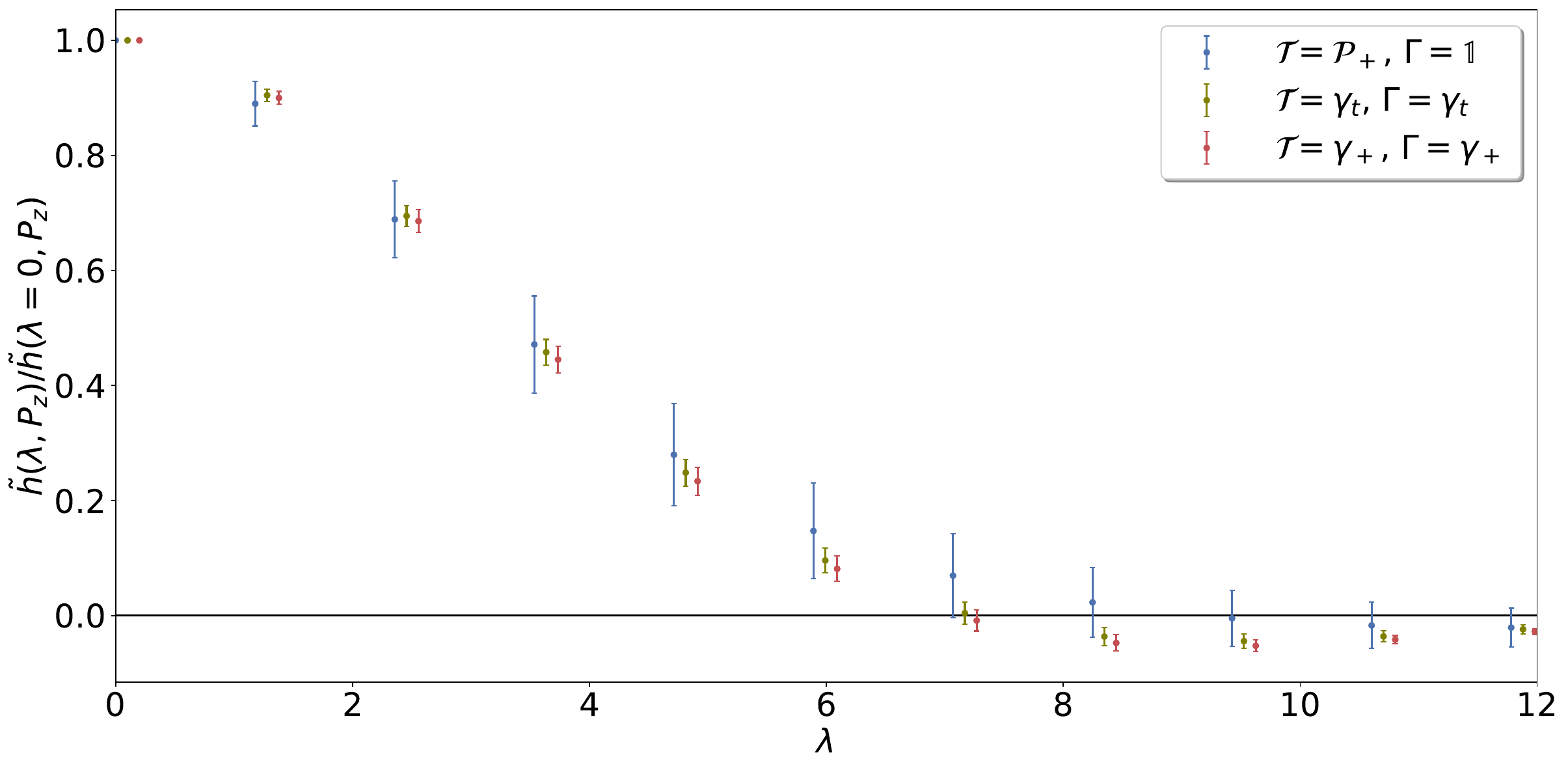}
    \caption{H102}
\end{subfigure}
\hfill
\begin{subfigure}{0.78\textwidth}
    \centering
    \includegraphics[width=\linewidth]{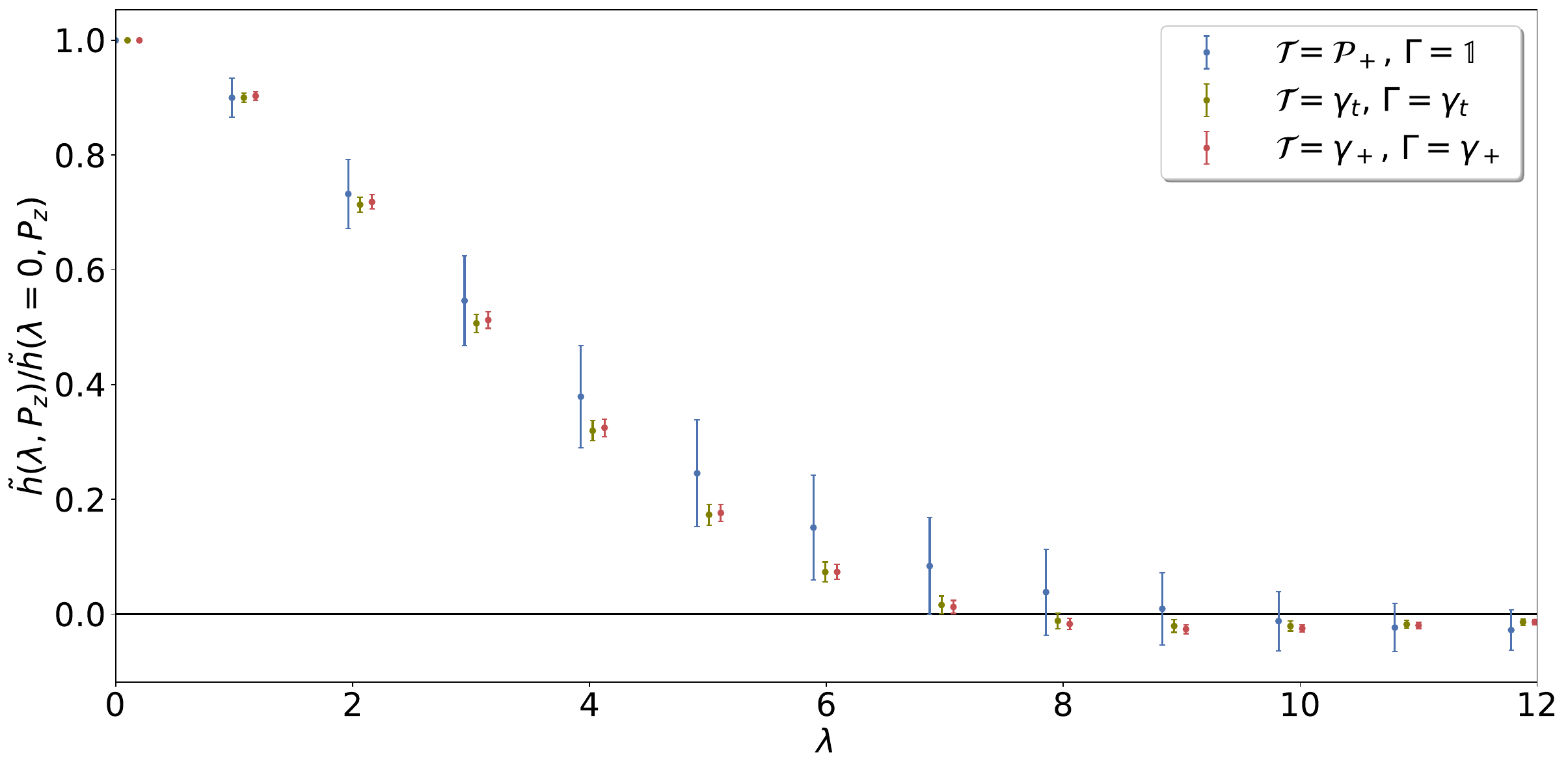}
    \caption{S400}
\end{subfigure}
\hfill
\begin{subfigure}{0.78\textwidth}
    \centering
    \includegraphics[width=\linewidth]{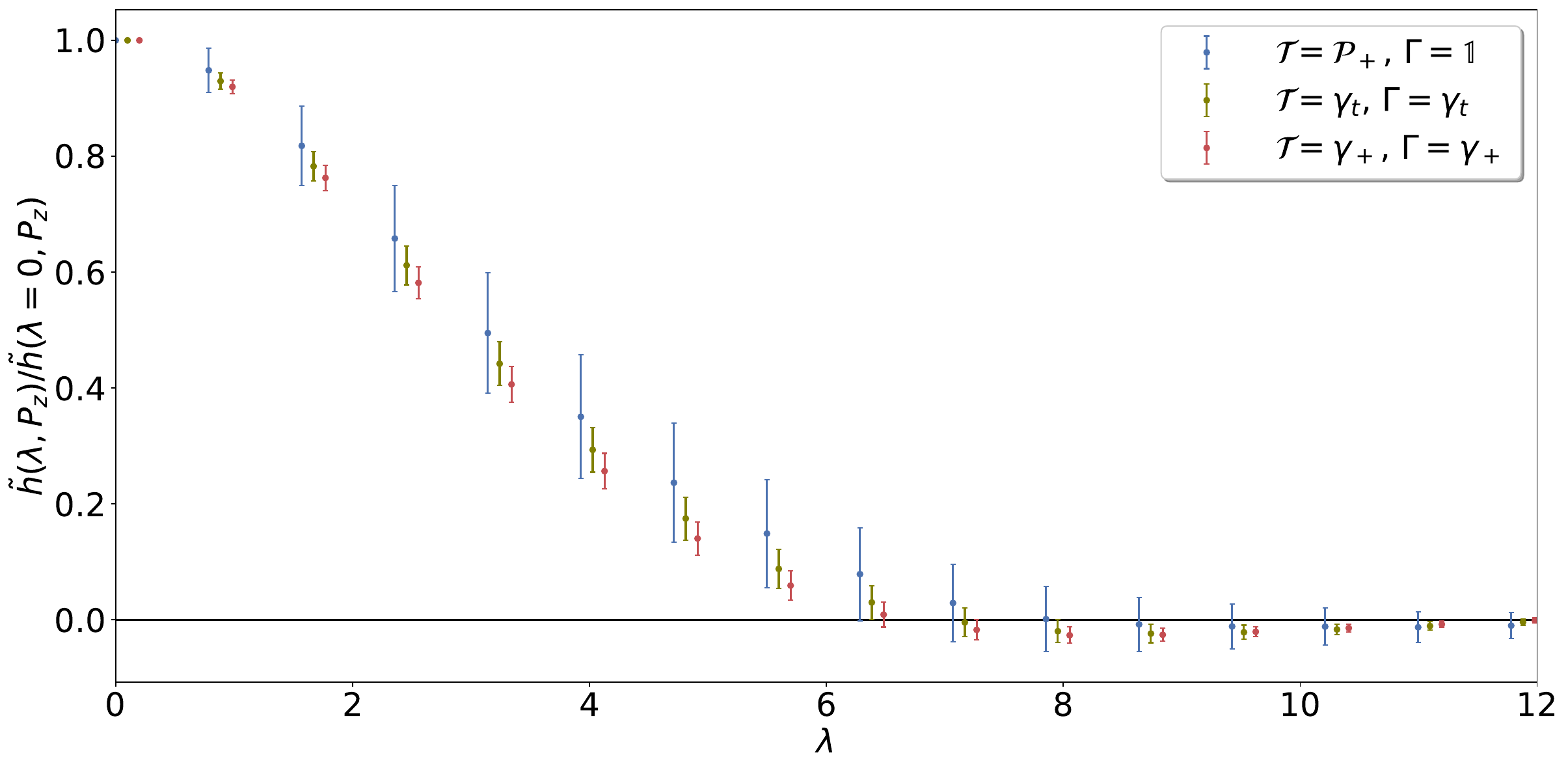}
    \caption{N203}
\end{subfigure}

\caption{Comparison of normalized bare matrix elements from different interpolators on H102, S400, and N203 ensembles (left to right) at momenta close to $P_z=2.5\,\mathrm{GeV}$. Data points have been offset horizontally to improve visibility.}
\label{fig:z_multistate_comparison}
\end{figure}

We present the bare unpolarized nucleon quark isovector matrix elements w.r.t. $\lambda=zP_z$ at $P_z\approx2.5\,\mathrm{GeV}$ in Figure \ref{fig:z_multistate_comparison} for all three ensembles used in this work. All calculations were performed at the same level of statistics; see Table \ref{tab:3pt_params}. A clear improvement in SNR is evident when using the kinematically enhanced interpolators. Across the three lattice ensembles, we observe an $\mathcal{O}(10)$ reduction in variance for the kinematically enhanced interpolators compared to the conventional one.

\begin{figure}[htbp]
    \centering
    
    \begin{subfigure}{0.78\textwidth}
        \centering
        \includegraphics[width=\linewidth, trim = 6 10 40 35, clip]{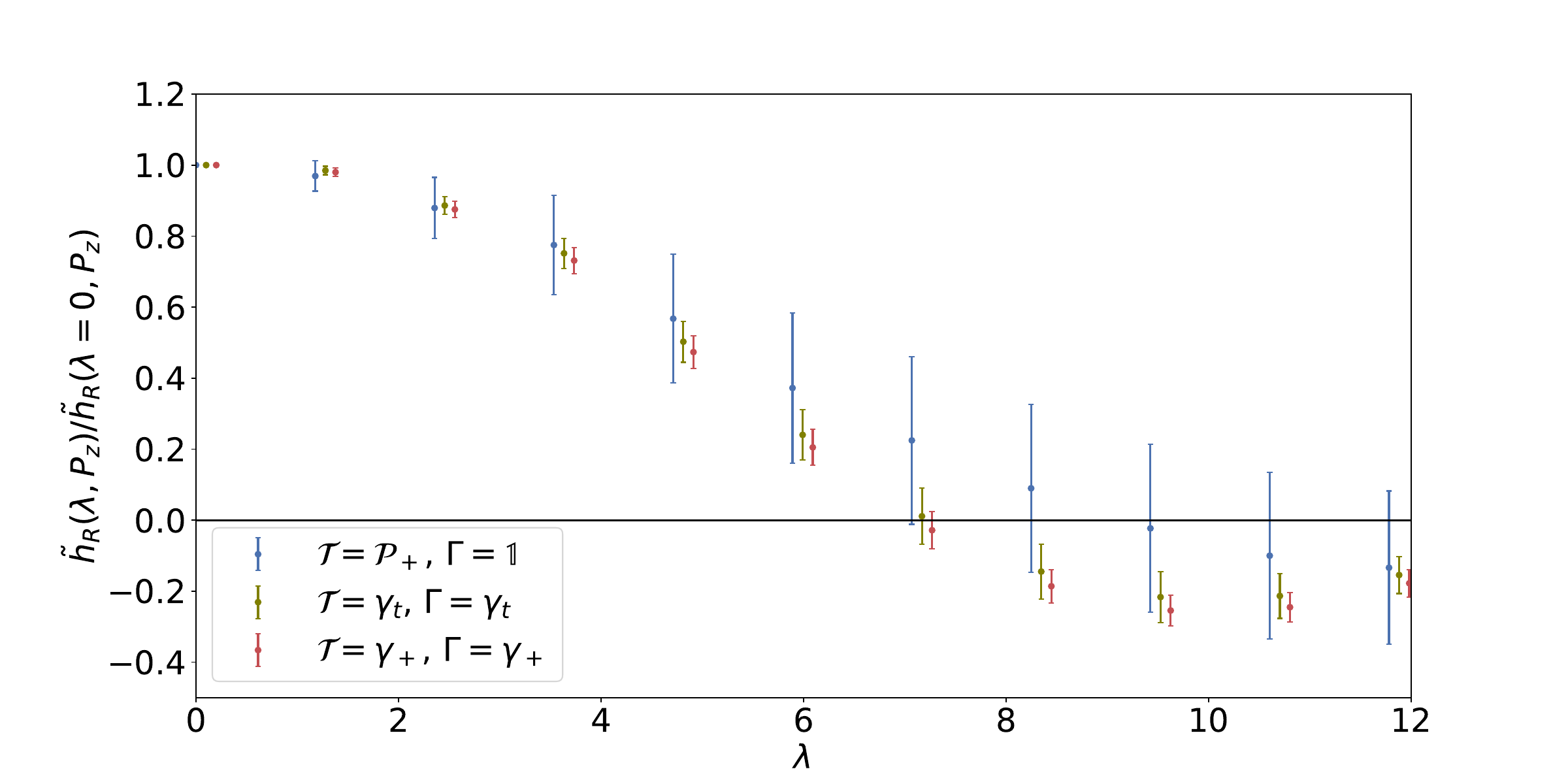}
        \caption{H102}
        \label{fig:ren_H102}
    \end{subfigure}
    \hfill
    \begin{subfigure}{0.78\textwidth}
        \centering
        \includegraphics[width=\linewidth, trim = 6 10 40 35, clip]{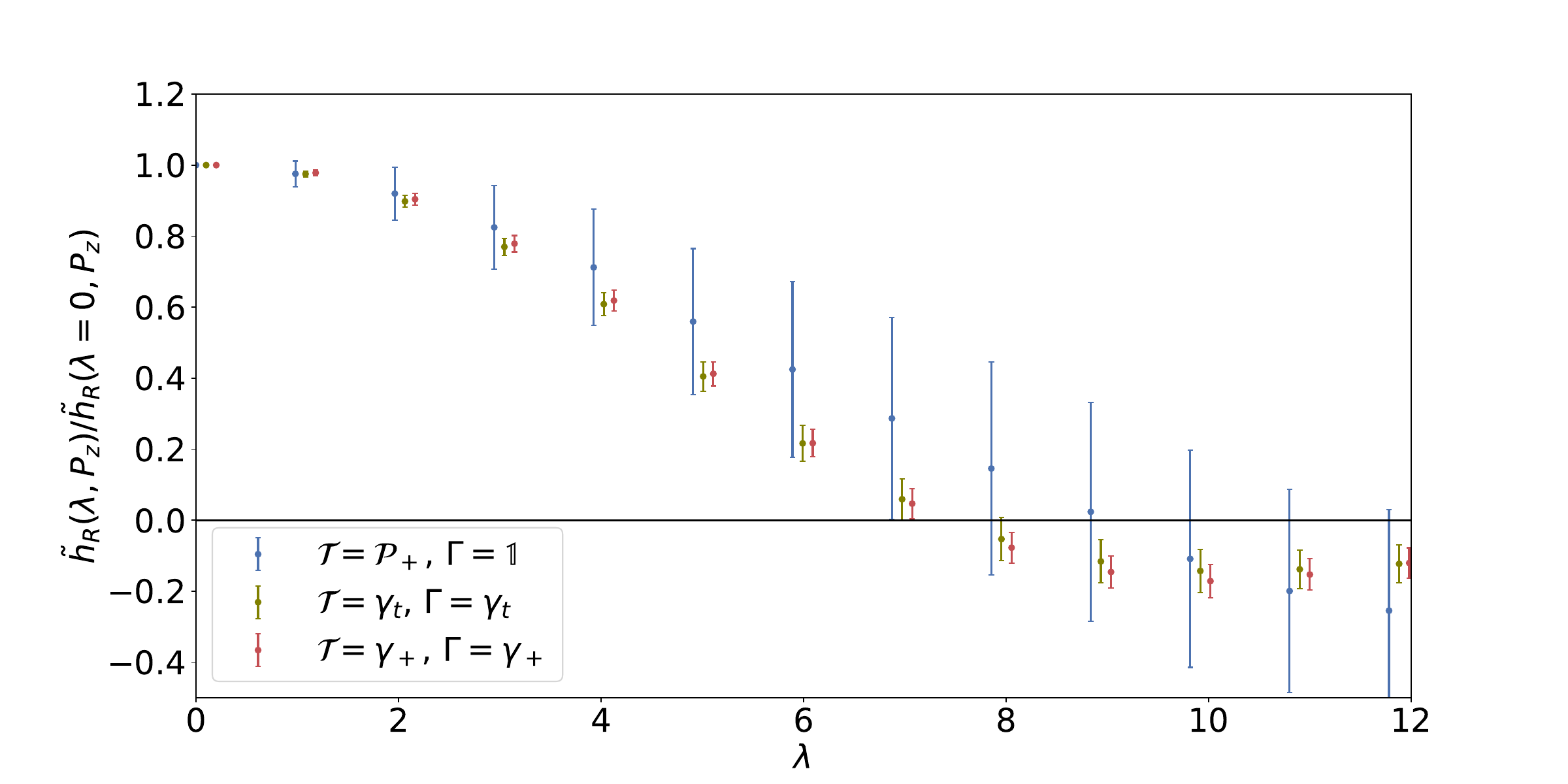}
        \caption{S400}
        \label{fig:ren_S400}
    \end{subfigure}
    \hfill
    \begin{subfigure}{0.78\textwidth}
        \centering
        \includegraphics[width=\linewidth, trim = 6 10 40 35, clip]{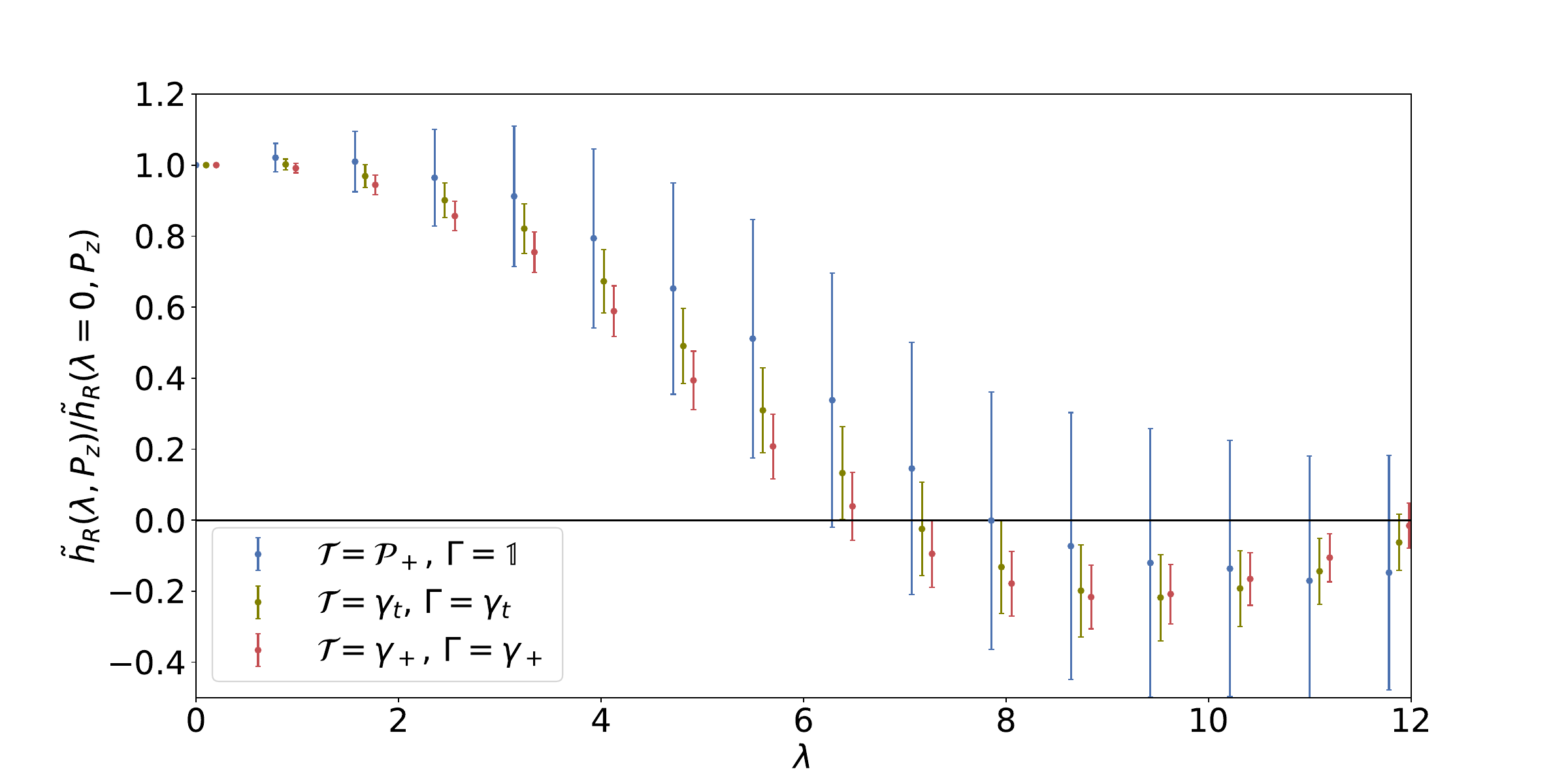}
        \caption{N203}
        \label{fig:ren_N203}
    \end{subfigure}

    \caption{Renormalized nucleon quark isovector matrix elements calculated on three ensembles for the different interpolator structures $\Gamma$ and $\mathcal{T}$. Datapoints at identical $\lambda$ have been slightly offset horizontally to improve visibility.}
    \label{fig:ren_ensemble_comparison}
\end{figure}

We renormalize the normalized matrix elements non-perturbatively using the hybrid renormalization scheme \cite{Ji:2020brr} as discussed in Subsection \ref{sec:renormalization}. We shall not elaborate on this further as it is not the focus of this work, instead referring the reader to Refs.~\cite{LatticeParton:2022xsd, Holligan:2024gpd, Niemiera:2025gluon, Holligan_2024, LatticePartonCollaborationLPC:2025vhd, Chen:2024rgi, LatticeParton:2024zko} for details. The required correlation functions in the rest frame were calculated with the conventional nucleon interpolator, as the kinematically enhanced ones are suboptimal choices at zero momentum. The ground state matrix elements at zero and large momenta were extracted, normalized by the local current results and their ratio taken subsequently. The linear divergence and renormalon term were extracted from a combined fit to the zero momentum ground state matrix elements on all three ensembles simultaneously~\cite{LatticePartonLPC:2021gpi,Zhang:2023bxs,Holligan:2023rex}. The results for the renormalized, unpolarized nucleon quasi-PDF obtained with different interpolators are shown in Figure \ref{fig:ren_ensemble_comparison}. As one can see, renormalization barely changes the enhancement we have observed in the bare matrix elements.

\begin{figure}[htbp]
\centering

\begin{subfigure}{0.78\textwidth}
    \centering
    \includegraphics[width=\linewidth, trim = 6 10 40 35, clip]{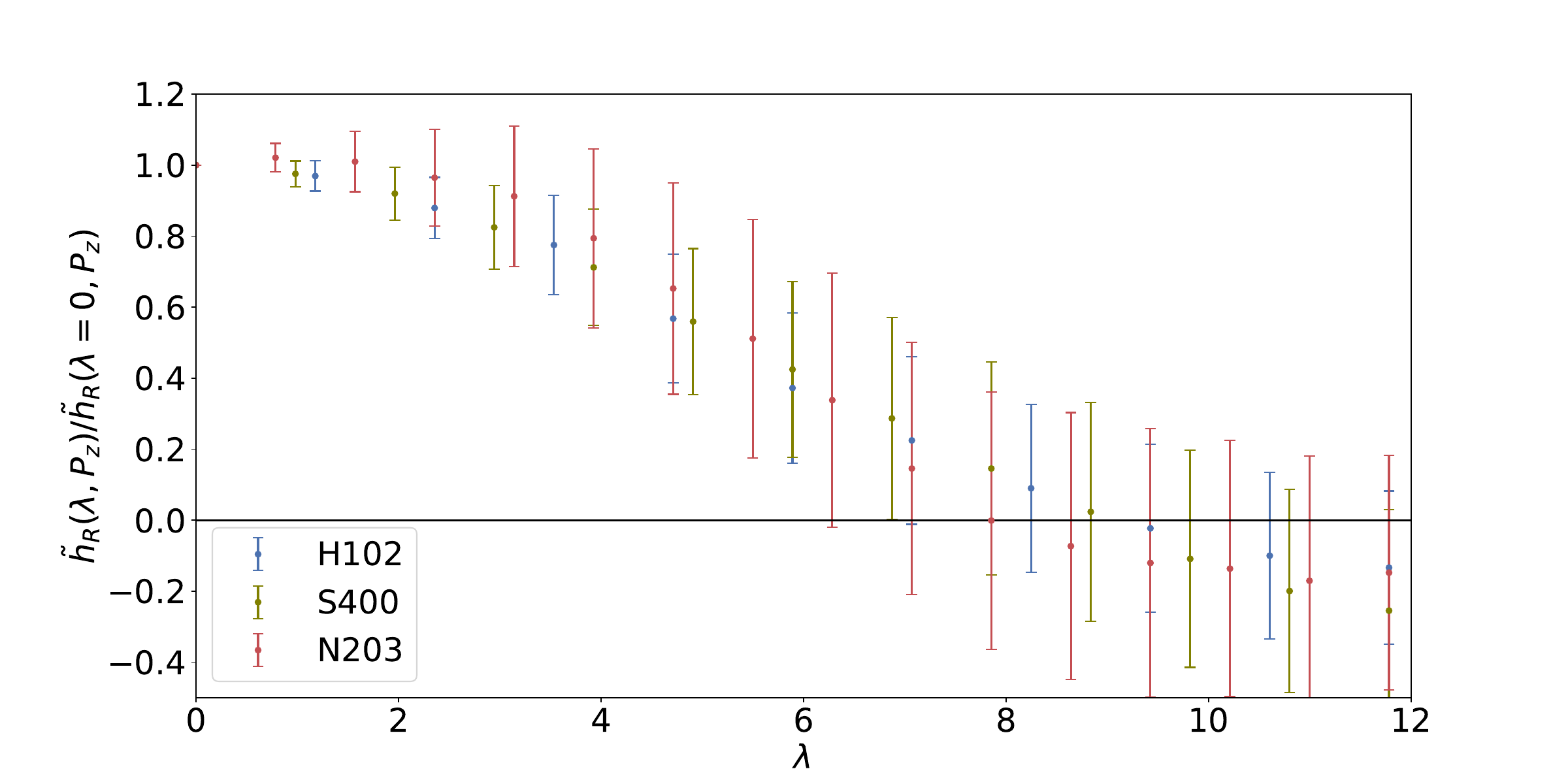}
    \caption{$(\mathcal{T},\Gamma)=(\mathcal{P}_+,\mathbbm{1})$}
\end{subfigure}
\hfill
\begin{subfigure}{0.78\textwidth}
    \centering
    \includegraphics[width=\linewidth, trim = 6 10 40 35, clip]{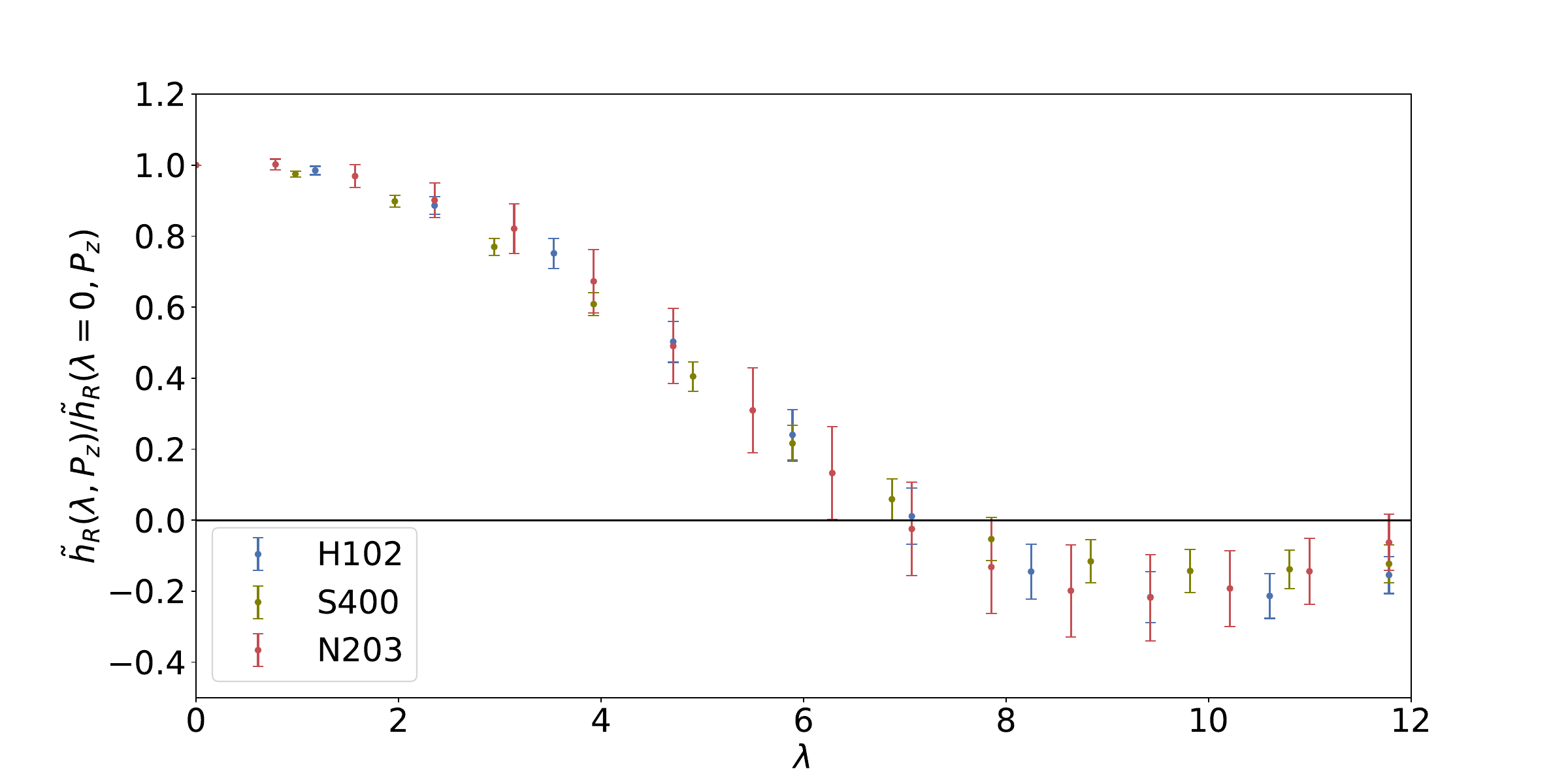}
    \caption{$(\mathcal{T},\Gamma)=(\gamma_t,\gamma_t)$}
\end{subfigure}
\hfill
\begin{subfigure}{0.78\textwidth}
    \centering
    \includegraphics[width=\linewidth, trim = 6 10 40 35, clip]{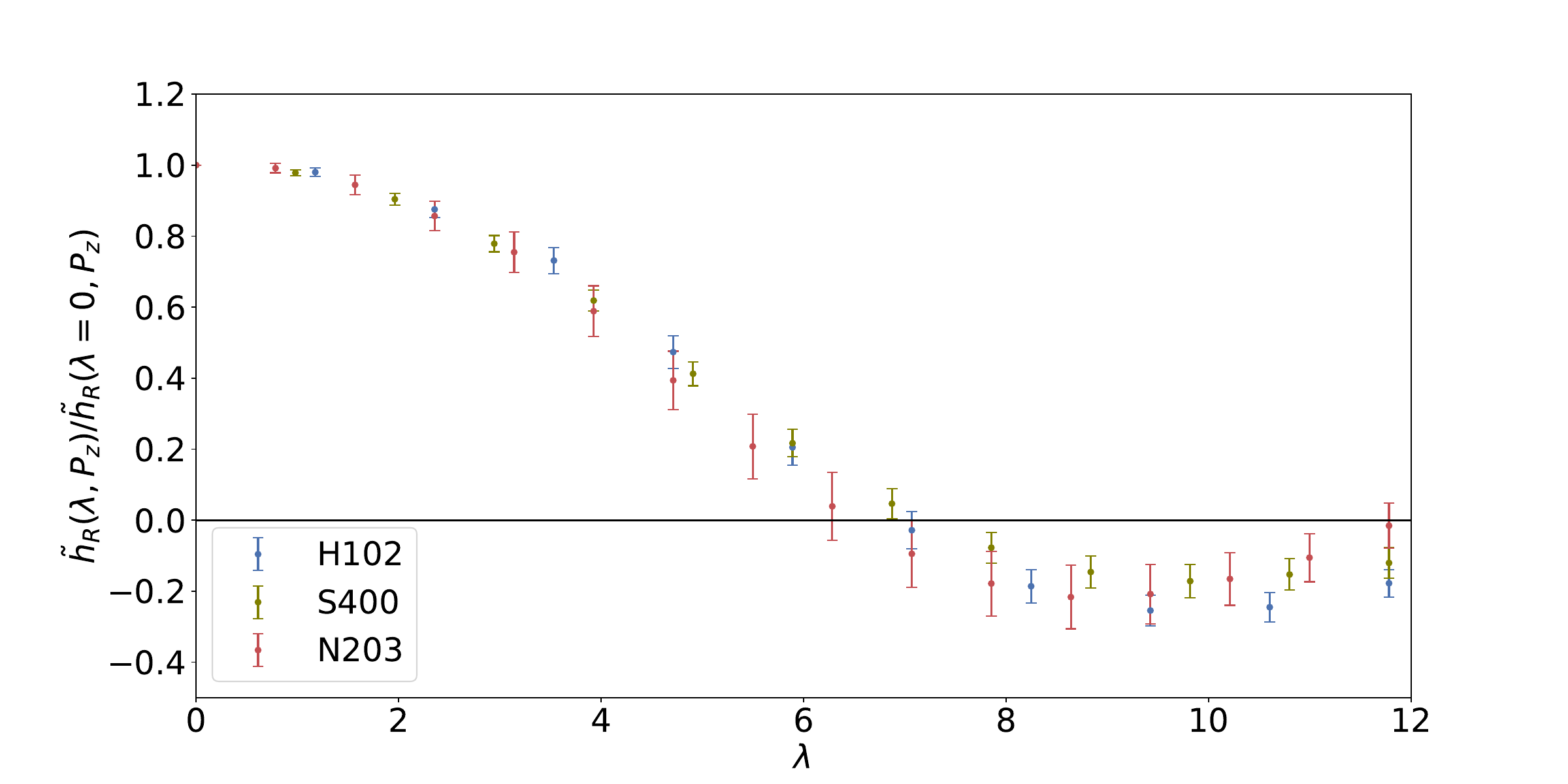}
    \caption{$(\mathcal{T},\Gamma)=(\gamma_+,\gamma_+)$}
\end{subfigure}

\caption{Renormalized nucleon quark isovector matrix elements at $P_z\approx2.5\,\mathrm{GeV}$ calculated on three ensembles of similar $m_\pi\approx350\,\mathrm{MeV}$ but different lattice spacing $a\in[0.064,0.086]\,\mathrm{fm}$ for the different interpolator structures $\Gamma$ and $\mathcal{T}$.}
\label{fig:three_subfigures}

\end{figure}

Finally, in Figure~\ref{fig:three_subfigures}, we compare the renormalized matrix elements at $P_z\sim 2.5$~GeV for the same interpolating operators across three lattice spacings: H102 ($a=0.085$~fm), S400 ($a=0.075$~fm), and N203 ($a=0.064$~fm). Although the exact values of $P_z$ differ slightly among the three ensembles, we observe no statistically significant dependence on $a$ within $1\sigma$ uncertainties for any of the interpolating operators considered, indicating that discretization effects remain small at such a large momentum. Since both discretization effects and LaMET power corrections depend on $P_z$, both the continuum and $P_z\to\infty$ limits must ultimately be taken to fully quantify the associated uncertainties. Our precise results with the new interpolating operators demonstrate the existence of a well-defined continuum limit, thereby providing a solid foundation for future precision LaMET calculations.

\section{Summary}
\label{section:summary}

Boosted nucleon matrix elements, relevant for parton structure, suffer from poor statistical precision at large momenta in lattice QCD. We investigate the use of kinematically enhanced interpolating operators for their calculation. Building on the proposal \cite{Zhang:2025hyo}, which demonstrated significant improvements for two-point correlation functions, we study their impact on gauge invariant nucleon quasi-PDF matrix elements on three CLS ensembles.

Our results show that the kinematically enhanced interpolators lead to an order-of-magnitude reduction in variance compared to conventional interpolators for nucleon matrix elements at $P_z\sim2.5$~GeV. We also observe that the enhancement factor increases toward finer lattice spacings. Moreover, the renormalized matrix elements remain consistent across three lattice spacings at similar $P_z$, indicating that discretization effects remain small even at such large momenta. These results place future precision LaMET calculations on a firmer foundation. Besides, our analysis also indicates that, when using the new interpolators, excited state contamination remains well-controlled and the extracted ground-state matrix elements are consistent with those obtained using conventional operators. Notably, the use of the projection operator $\mathcal{T}=\gamma_+$ together with $\Gamma=\gamma_+$ in the interpolator provides an additional factor-of-four gain in statistics, further reducing the variance of the matrix elements. 


Our findings demonstrate that kinematically enhanced interpolators provide a practical and robust method for improving the quality of lattice calculations involving boosted nucleons. Their implementation can significantly enhance the statistical reach of such studies without introducing additional systematic uncertainties, thereby enabling more ambitious nucleon structure calculations within the LaMET framework, while also benefiting other approaches and physical applications that require large hadron momenta.

In the future, we plan to explore momenta $P_z\gtrsim 3$~GeV, which should further improve the precision of LaMET calculations. However, without employing finer lattices, the ${\cal O}((aP_z)^n)$ discretization effects are expected to become increasingly important as $P_z$ grows. To mitigate these effects, one possible direction is to adopt the Coulomb-gauge method~\cite{Gao:2023lny,Zhao:2023ptv,Gao:2026hix}, where quasi-PDFs are constructed from gauge-fixed quark correlators without Wilson lines. In this framework, the hadron momentum $\vec{P}$ can be chosen along a diagonal direction, increasing its magnitude by up to a factor of $\sqrt{3}$ without enhancing discretization effects~\cite{Gao:2023lny}. Moreover, the Coulomb-gauge method itself offers exponentially improved SNRs for long-range matrix elements, which can be combined with kinematically enhanced interpolators to further suppress statistical uncertainties.


\acknowledgments

DR and AS are supported by the Deutsche Forschungsgemeinschaft  (DFG) grant SCHA 458/23, project number 493441321.
TS is supported by the Deutsche Forschungsgemeinschaft (DFG) Research Unit FOR 2926, "Next Generation pQCD for Hadron Structure: Preparing for the EIC", project number 40824754. 
RZ is supported by the U.S. Department of Energy, Office of Science, Office of Nuclear Physics under grant Contract No.~DE-SC0011090 and  DOE Quark-Gluon Tomography (QGT) Topical Collaboration under award No.~DE-SC0023646.
YZ is supported by the U.S. Department of Energy, Office of Science, Office of Nuclear Physics through Contract No.~DE-AC02-06CH11357, and the Early Career Award through Contract No.~DE-SCL0000017.

The authors gratefully acknowledge the scientific support and HPC resources provided by the Erlangen National High Performance Computing Center (NHR@FAU) of the Friedrich-Alexander-Universität Erlangen-Nürnberg (FAU) under the NHR project b164da. NHR funding is provided by federal and Bavarian state authorities. NHR@FAU hardware is partially funded by the German Research Foundation (DFG) – 440719683.
The authors gratefully acknowledge the Gauss Centre for Supercomputing e.V. (www.gauss-centre.eu) for funding this project by providing computing time through the John von Neumann Institute for Computing (NIC) on the GCS Supercomputer JUPITER at Jülich Supercomputing Centre (JSC).

We acknowledge the CLS effort for generating the $n_f = 2 + 1$ ensembles in
\cite{Bruno:2014jqa}, three of which were used for this work. 

\appendix

\section{Bare $C_{3pt}$ improvement}

\begin{figure}[htbp]
    \centering
    \begin{subfigure}{1.0 \textwidth}
        \centering
        \includegraphics[width=\linewidth]{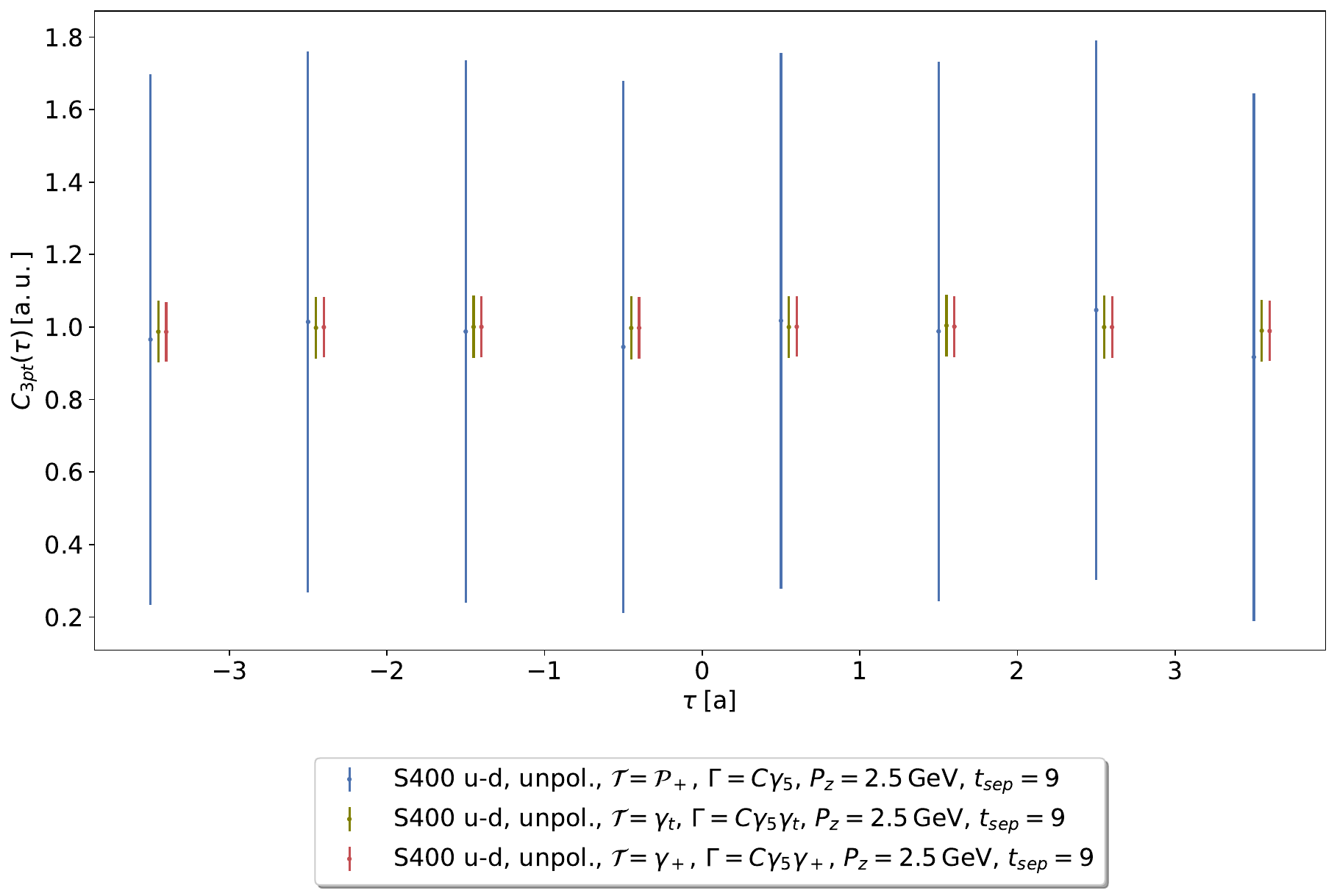}
        \caption{$t_{\mathrm{sep}} = 9$}
        \label{fig:3pt_t9}
    \end{subfigure}
    
    \caption{Bare, normalized unpolarized nucleon isovector charge three-point function for different interpolators, obtained from 400 measurements on the S400 ensemble at $|P_z|=2.5\,\mathrm{GeV}$. $\tau$ is the insertion time relative to $t_\mathrm{sep}/2$}
    \label{fig:3pt_tsep}
\end{figure}

An improvement similar to the one shown in Figure \ref{fig:2pt_all}
is observed for the three-point function $C_{3pt}$, presented in Figure \ref{fig:3pt_tsep}. As the numerical values of the overlap factors, which later will be canceled in the $C_{3pt}/C_{2pt}$, differ vastly, we divided the data points by the weighted average of the central points for better comparison of the error bars. Clearly, the SNR is improved when using the kinematically enhanced interpolators, particularly at larger separation times. 




\phantomsection
\addcontentsline{toc}{section}{References}

\bibliographystyle{JHEP}
\bibliography{main.bib}

\end{document}